\documentclass[prd,aps,twocolumn,a4paper,floatfix,nofootinbib]{revtex4-1}

\usepackage[utf8]{inputenc}
\usepackage{graphicx,psfrag}
\usepackage{mathrsfs}
\usepackage{amsmath,amsfonts,amssymb}
\usepackage{multirow}
\usepackage{comment}
\usepackage{xcolor}
\usepackage{enumerate}
\usepackage{hyperref}
\hypersetup{
    colorlinks = true,
    linkcolor = {blue},
    citecolor = {blue},
    urlcolor = {blue},
    linkbordercolor = {white},
    }

\newcommand{\be}{\begin{equation}}
\newcommand{\ee}{\end{equation}}
\newcommand{\bea}{\begin{eqnarray}}
\newcommand{\eea}{\end{eqnarray}}
\newcommand{\bel}{\begin{align}}
\newcommand{\eel}{\end{align}}

\newcommand{\Suu}{SLy$^{(\uparrow \uparrow)}$}
\newcommand{\Sdd}{SLy$^{(\downarrow \downarrow)}$}
\newcommand{\Swe}{SLy$^{(\leftarrow \rightarrow)}$}
\newcommand{\Snw}{SLy$^{(\nwarrow \nearrow)}$}
\newcommand{\Ssw}{SLy$^{(\swarrow \searrow)}$}
\newcommand{\Snee}{SLy$^{(\nearrow \nearrow)}$}
\newcommand{\Ssee}{SLy$^{(\searrow \searrow)}$}
\newcommand{\Soo}{SLy$^{(00)}$}

\def\GMc2{{\rm G M_{\odot} c^{-2}}}

\usepackage{color}

\definecolor{cyan}{rgb}{0,0.9,0.9}
\definecolor{orange}{rgb}{0.9,0.5,0}
\definecolor{magenta}{rgb}{1,0,1}
\definecolor{purple}{rgb}{0.8,0.4,0.8}
\definecolor{gray}{rgb}{0.8242,0.8242,0.8242}
\definecolor{mgreen}{rgb}{0.1,0.8,0.1}

\usepackage[normalem]{ulem}

\begin{document}

\title{Gravitational waves and mass ejecta from binary neutron star mergers:
Effect of the spin orientation}

\author{Swami Vivekanandji \surname{Chaurasia}$^{1,2}$}
\author{Tim \surname{Dietrich}$^{3,4}$}
\author{Maximiliano \surname{Ujevic}$^5$}
\author{Kai Hendriks$^{1,6,9}$}
\author{Reetika \surname{Dudi}$^{1,7}$}
\author{Francesco Maria \surname{Fabbri}$^1$}
\author{Wolfgang \surname{Tichy}$^8$}
\author{Bernd \surname{Br\"ugmann}$^1$}

\affiliation{${}^1$Theoretical Physics Institute, University of Jena, 07743 Jena, Germany}
\affiliation{${}^2$The Oskar Klein Centre, Department of Astronomy, Stockholm University, AlbaNova, SE-10691 Stockholm,
Sweden}
\affiliation{${}^3$Institut f\"{u}r Physik und Astronomie, Universit\"{a}t Potsdam, Haus 28, Karl-Liebknecht-Strasse 24/25, 14476, Potsdam, Germany}
\affiliation{${}^4$Nikhef, Science Park 105, 1098 XG Amsterdam, Netherlands}
\affiliation{${}^5$Centro de Ci\^encias Naturais e Humanas, Universidade Federal do ABC, 09210-170, Santo Andr\'e, S\~ao Paulo, Brazil}
\affiliation{${}^6$Department of Astrophysics/IMAPP, Radboud University, P.O. Box 9010, 6500 GL Nijmegen, Netherlands}
\affiliation{${}^7$Max Planck Institute for Gravitational Physics, Albert Einstein Institute, D-14476 Golm, Germany}
\affiliation{${}^8$Department of Physics, Florida Atlantic University, Boca Raton, FL 33431 USA}
\affiliation{${}^9$Maastricht Science Programme, Faculty of Science and Engineering, Maastricht University, P.O. Box 616, 6200 MD Maastricht, Netherlands}

\date{\today}

\begin{abstract}
We continue our study of the binary neutron star parameter space by investigating the effect 
of the spin orientation on the dynamics, gravitational wave emission, and mass ejection during the 
binary neutron star coalescence. 
We simulate seven different configurations using multiple resolutions to allow a reasonable 
error assessment. 
Due to the particular choice of the setups, 
five configurations show precession effects, 
from which two show a precession (``wobbling'') 
of the orbital plane, while three show a ``bobbing'' motion, i.e., the orbital angular momentum 
does not precess, while the orbital plane moves along the orbital angular momentum axis. 
Considering the ejection of mass, we find that precessing systems can have an anisotropic 
mass ejection, which could lead to a final remnant kick of $\sim 40 \rm km/s$ 
for the studied systems. Furthermore, for the chosen configurations, antialigned spins lead to 
larger mass ejecta than aligned spins, so that brighter electromagnetic counterparts could 
be expected for these configurations. 
Finally, we compare our simulations with the precessing, tidal waveform approximant 
\texttt{IMRPhenomPv2\_NRTidalv2} and find good agreement between the approximant 
and our numerical relativity waveforms with phase differences below 
1.2 rad accumulated over the last $\sim$ 16 gravitational wave cycles. 
\end{abstract}


\maketitle

\section{Introduction}
\label{sec:intro}

The first coincidence detection of gravitational waves (GWs) and electromagnetic (EM) waves 
originating from the same astrophysical source, the binary neutron star (BNS) merger 
GW170817, inaugurated a new era in multimessenger 
astronomy~\cite{TheLIGOScientific:2017qsa,Monitor:2017mdv}. 
Already this first BNS detection provided important scientific insights, e.g., 
it allowed for a new and independent measurement of the Hubble 
constant (e.g.,~\cite{Abbott:2017xzu,Coughlin:2019vtv}),  
it proved that NS mergers are a source of r-process elements
(e.g.,~\cite{Cowperthwaite:2017dyu,Smartt:2017fuw,Kasliwal:2017ngb,Kasen:2017sxr,Watson:2019xjv}), 
and it placed constraints on the equation of state (EOS) of cold matter at supranuclear densities 
(e.g.,~\cite{TheLIGOScientific:2017qsa,Abbott:2018wiz,Radice:2018ozg,Coughlin:2018fis,Capano:2019eae,Dietrich:2020lps}).
In addition, the increasing number of potential binary neutron star
candidates and the second confirmed detection
of a binary neutron star merger, GW190425~\cite{Abbott:2020uma},
suggest that many more systems will be detected in the near future.

For a correct analysis and interpretation of the observed signals, 
one has to relate the measured data with theoretical predictions. 
With respect to GW astronomy, this can be done by 
correlating the signal with a waveform model maximizing their agreement, 
e.g.,~\cite{Veitch:2014wba}.
Considering EM astronomy, one needs to relate the observed 
properties of the signals (spectra and light curves) 
with the theoretical predictions of EM transients, which are connected to 
the material outflow and evolution during the last stages of the binary dynamics, 
e.g.~\cite{Cowperthwaite:2017dyu,Smartt:2017fuw,Kasliwal:2017ngb,
Kasen:2017sxr,Watson:2019xjv,Coughlin:2018fis}. \\

To be prepared for future detections of BNS systems with various intrinsic parameters, one has to cover the entire parameter space; i.e., one has to vary systematically the individual masses, the neutron stars (NSs) spins. In addition, our missing knowledge about the exact EOS adds an additional free parameter that we need to vary in our studies.
In this article, we will focus on the effect of intrinsic NS spin on the BNS coalescence. 

Although pulsar observations of BNS systems suggest 
that most NSs have small spins, e.g.,
\cite{Kiziltan:2013oja,Lattimer:2012nd}, this
conclusion is based on a small selected set of observed binaries. 
Observations of isolated NSs or NSs in binary systems other than BNSs
show that NSs can rotate fast; e.g., 
PSR J1807$-$2500B has a rotation frequency of $239$Hz~\cite{Lorimer:2008se,Lattimer:2012nd}. 

Similar to the uncertainty in the spin magnitude, 
the orientation of spins in BNS systems is also 
highly uncertain and unknown. Misaligned spins can be caused by 
the supernova explosions of the progenitor stars. 
A possible realignment of the spin with the orbital angular momentum 
due to accretion is only possible for the more massive NS, 
but not for the secondary star; e.g., 
for PSRJ0737-3039B the angle between the spin and the orbital angular momentum 
is $\approx 130^\circ$~\cite{Farr:2011gs}. 
In addition, for BNS systems formed due to dynamical capture, there is no reason to have aligned spins at all and one can expect that spins will be isotropically distributed.
Consequently, further investigations of the effect of the spin orientation 
are required.

We will present a detailed numerical relativity study for various precessing systems. 
We point out that, in most numerical relativity (NR) studies, spins have been neglected or 
have been treated unrealistically by assuming that the stars are tidally locked.
Only in the last few years, NR groups performed spinning NS simulations
dropping the corotational assumption. 
The only NR simulations in which the Einstein constraint equations 
and also the equations of general relativistic
hydrodynamics are solved for configurations 
in which the individual NSs are spinning, are presented 
in~\cite{Bernuzzi:2013rza,Dietrich:2015pxa,Dietrich:2016lyp,
Dietrich:2018upm,Most:2019pac,Tsokaros:2019anx,East:2019lbk}. 
With respect to precession, the list of studies is even 
shorter~\cite{Dietrich:2015pxa,Tacik:2015tja,Dietrich:2017xqb}.
Ref.~\cite{Dietrich:2015pxa} performed a preliminary study
for one precessing, one spin aligned, and one nonspinning configuration 
employing only low resolution grid setups. 
A precessing inspiral has also been shown 
in~\cite{Tacik:2015tja}, but the merger and postmerger parts have been excluded. 
Finally, \cite{Dietrich:2017xqb} performed a more
systematic study for two unequal-mass, precessing NS systems. 
In total, the entire NR community has studied less than five precessing configurations until now. 
To overcome this shortage, we study several equal-mass BNS configurations for various spin orientations. 
Each configuration is evolved with
four different resolutions.

The article is structured as follows: Sec.~\ref{sec:simu}
describes the numerical methods that we employ and the 
configurations that we study. 
In Sec.~\ref{sec:dyn} we provide a first discussion about the coalescence
by focusing on the energetics and the properties of the
merger remnant. In Sec.~\ref{sec:ejectaandkicks} 
we discuss the mass ejection and kick estimates for
the studied configurations. In Sec.~\ref{sec:GWs}
we study the emitted GW signal by analyzing the phase
evolution for the different setups, compare the waveforms with GW approximants, 
and comment on the postmerger frequencies.
We conclude in Sec.~\ref{sec:summary}.
For completeness, we give important expressions for the computation of
radiated energy, angular momentum, and linear momentum in Appendix~\ref{app:EJ_formulas}
and discuss in Appendix~\ref{app:accuracy} the accuracy of our NR simulations.

\section{Methods and Configurations}
\label{sec:simu}

\subsection{Numerical methods}

\begin{table*}[t]
\caption{BNS configurations. 
The first column gives the configuration name.
The next five columns provide the physical properties of the individual stars: the gravitational masses of the individual stars $M^{A,B}$, the 
baryonic masses of
the individual stars $M _b ^{A,B}$, the stars' dimensionless spins magnitude
$\chi^{A,B}$ and their orientations $\hat{\chi}^A$ and 
$\hat{\chi}^B$.
The last six columns give the mass-weighted effective spin $\chi _\text{eff}$, the effective
spin-precession parameter $\chi _p$, the residual eccentricity $e$, the initial GW
frequency $M\omega^0 _{22}$, the Arnowitt-Deser-Misner (ADM) mass $M_\text{ADM}$, and the ADM angular
momentum $J_\text{ADM}$. The configurations were evolved with the resolutions of Table~\ref{tab:grid}.}
\label{tab:config} 
\begin{tabular}{cccccccccccc}
\toprule
Name & $M^{A,B}$ & $M^{A,B}_b$ & $\chi^{A,B}$ & $\hat{\chi}^A$ & $\hat{\chi}^B$ & $\chi _\text{eff}$ & $\chi _\text{p}$ & $e$ & $M\omega ^0 _{22}$ & $M_\text{ADM}$ & $J_\text{ADM}$ \\ 
\hline
\Suu & 1.3505 & 1.4946 & 0.0955 & (0,0,1) & (0,0,1) & 0.0955 & 0 & 0.00753 & 0.03405 & 2.6799 & 8.1939 \\
\Snw & 1.3505 & 1.4946 & 0.0956 & $\frac{(-1,0,1)}{\sqrt{2}}$ & $\frac{(1,0,1)}{\sqrt{2}}$ & 0.0676 &
0.0676 & 0.00793 & 0.03406 & 2.6799 & 8.0993 \\
\Snee & 1.3505 & 1.4946 & 0.0955 & $\frac{(1,0,1)}{\sqrt{2}}$ & $\frac{(1,0,1)}{\sqrt{2}}$ & 0.0675 & 0.0676 & 0.00813 & 0.03406 & 2.6799 & 8.1020 \\
\Swe & 1.3505 & 1.4946 & 0.0955 & (-1,0,0) & (1,0,0) & 0 & 0.0955 & 0.00922 & 0.03408 & 2.6799 & 7.8712 \\
\Ssw & 1.3505 & 1.4946 & 0.0956 & $\frac{(-1,0,-1)}{\sqrt{2}}$ & $\frac{(1,0,-1)}{\sqrt{2}}$ & -0.0676 & 0.0676 & 0.01083 & 0.03411 & 2.6799 & 7.6437 \\
\Ssee & 1.3505 & 1.4946 & 0.0956 & $\frac{(1,0,-1)}{\sqrt{2}}$ & $\frac{(1,0,-1)}{\sqrt{2}}$ & -0.0676 & 0.0676 & 0.01194 & 0.03409 & 2.6799 & 7.6437 \\
\Sdd & 1.3505 & 1.4946 & 0.0955 & (0,0,-1) & (0,0,-1) & -0.0955 & 0 & 0.01197 & 0.03411 & 2.6799 & 7.5484 \\\hline 
\hline
\end{tabular} 
\end{table*}

\subsubsection{Initial data construction}

The initial data for the setups studied
in this article are obtained with
the pseudospectral \texttt{SGRID}
code~\cite{Tichy:2006qn,Tichy:2009yr,Tichy:2009zr,Dietrich:2015pxa}.
Quasiequilibrium configurations of NSs
with arbitrary spins and different
EOSs~\cite{Dietrich:2015pxa} can be obtained with
\texttt{SGRID}\footnote{This project started before the upgraded \texttt{SGRID} version 
presented in~\cite{Tichy:2019ouu} was available, 
so that we have used the previous \texttt{SGRID} version 
of~\cite{Dietrich:2015pxa} and therefore could not explore higher
spins possible with the upgraded version.},
 which employs the conformal thin sandwich
formalism~\cite{Wilson:1995uh,Wilson:1996ty,York:1998hy}
in addition to the constant rotational velocity
approach~\cite{Tichy:2011gw,Tichy:2012rp,Tichy:2016vmv} to describe
the rotation state of the NSs.
Although \texttt{SGRID} can construct eccentricity reduced initial data, 
we do not perform any kind of eccentricity reduction
to reduce computational costs.
Moreover, the residual eccentricities for our
quasiequilibrium setups are reasonably small ($\lesssim 10^{-2}$) 
for our present analysis~\cite{Dietrich:2015pxa};
see Table~\ref{tab:config}.

The computational domain of \texttt{SGRID} is divided into six patches
(Fig.~1 of \cite{Dietrich:2015pxa}) that includes spatial infinity, 
which allows imposing exact
boundary conditions. We employ $n_A = n_B=28$,
$n_\varphi=8$, $n_{\rm Cart} = 24$ points for
the spectral grid; cf.~\cite{Tichy:2006qn,Tichy:2009yr,Tichy:2009zr,Dietrich:2015pxa,Tichy:2019ouu}
for further details.

\begin{table}[t]
\caption{Grid configurations. The columns refer to: the resolution name, 
the number of levels $L$, the number of moving box levels $L_{\rm mv}$, 
the number of points in the nonmoving boxes $n$, the number of points 
in the moving boxes $n_{\rm mv}$, the grid spacing in the finest level 
$h_6$ covering the NS diameter, the grid spacing in the coarsest level 
$h_0$, and the outer boundary position $R_0$. The grid spacing and the 
outer boundary position are given in units of $M_{\odot}$}
\label{tab:grid} 
\centering
\begin{tabular}{cccccccc}
\toprule
Name & $L$ & $L_{\rm mv}$ & $n$ & $n_{\rm mv}$ & $h_6$ & $h_0$ & $R_0$ 
\\ \hline
R1 & 7 & 3 & 192 & 64 & 0.246 & 15.744 & 1511.4 \\
R2 & 7 & 3 & 288 & 96 & 0.164 & 10.496 & 1511.4 \\
R3 & 7 & 3 & 384 & 128 & 0.123 & 7.872 & 1511.4 \\
R4 & 7 & 3 & 480 & 160 & 0.0984 & 6.2976 & 1511.4 \\ \hline \hline
\end{tabular}
\end{table}

\subsubsection{Dynamical evolutions}

The constructed initial data are evolved
with the \texttt{BAM} code~\cite{Brugmann:2008zz,Thierfelder:2011yi,Dietrich:2015iva,
Bernuzzi:2016pie}, utilizing the
Z4c formulation of the Einstein equations
for the evolution system~\cite{Bernuzzi:2009ex,Hilditch:2012fp} 
together with the (1+log)-lapse and gamma-driver-shift 
conditions~\cite{Bona:1994a,Alcubierre:2002kk,vanMeter:2006vi}.
The numerical fluxes for the general relativistic hydrodynamics 
system are constructed with a flux-splitting approach based on 
the local Lax-Friedrich (LLF) flux. We perform the flux reconstruction with 
a fifth-order WENOZ algorithm~\cite{Borges:2008a} on 
the characteristic fields~\cite{Jiang:1996,Suresh:1997,Mignone:2010br} 
to obtain high-order convergence~\cite{Bernuzzi:2016pie}.
For low density regions and around the moment of merger,
we switch to a primitive reconstruction scheme that is more stable
but less accurate i.e., from a higher-order LLF scheme that uses the characteristic fields 
to a second-order LLF scheme that simply uses the primitive variables~\cite{Bernuzzi:2016pie}.
A piecewise-polytropic form of the EOS
approximation is used for the SLy EOS~\cite{Read:2009yp}.
Additionally, thermal effects to
the EOS are added by a thermal pressure following an ideal
gas contribution i.e., by adding an additional thermal pressure of the form $p_{\rm th} = \rho \epsilon
(\Gamma_\text{th}-1)$ with $\Gamma_{\rm th}=1.75$;
see~\cite{Bauswein:2010dn}.

The method of lines is used for the time integration combined with an explicit fourth-order Runge-Kutta integrator.
Furthermore, the time stepping utilizes the Berger-Collela
scheme, enforcing mass conservation across the
refinement boundaries~\cite{Berger:1984zza,Dietrich:2015iva}.

The computational domain is divided into a hierarchy of cell 
centered nested Cartesian grids with refinement factor of $2$.
Each level has one or more Cartesian grids with constant grid spacing
$h_l$ and $n$ (or $n^{\rm mv}$) points per direction.
Some of the refinement levels $l > l^{\rm mv}$
can be dynamically moved and adapted during the time evolution
according to the technique of ``moving boxes''. In this article, 
we set $l^{\rm mv} = 3$.

Since we are interested in spin and precession effects,
we cannot enforce any additional symmetry and evolve the
full 3D grid. This increases the computational costs
by a factor of $2$ compared to most of our
past studies where we employed bitant symmetry.
In order to have compatible simulations even the
spin-aligned and antialigned setups that are not
expected to show any precession are evolved without
imposing any symmetry.
Details about the different grid configurations 
employed in this work are given in Tab.~\ref{tab:grid};
the grid configurations are labeled as R1, R2, R3, R4, 
ordered by increasing resolution.

\subsection{Configurations}

In this article we study equal-mass systems with NSs
at an initial proper separation of $\sim$ 56 km and
having fixed rest masses (baryonic masses) of
$M^{A,B} _b = 1.4946M_{\odot}$.
The gravitational masses for the NSs in isolation are $M^{A,B} \simeq 1.35M_{\odot}$,
leading to a binary mass of $M \simeq 2.70M_{\odot}$,
see details in Tab.~\ref{tab:config}.
The individual stars are spinning and have dimensionless spins
$\chi^A=\chi^B\approx0.096$ which corresponds to $\sim 190$ Hz
for the SLy EOS used in this study.
The simulated configurations
differ in their spin orientation with respect to the orbital
angular momentum direction of the system.
We note that a setup in which only one star has a non-negligible spin
might be astrophysically better motivated. However, our
current study is pedagogically motivated. Moreover, we expect to maximize
the effects of misaligned-spin from the chosen configurations.
Keeping the systems symmetric we expect to better
disentangle the effect of misaligned-spins and have better
quantitative comparisons among the simulated setups.
In Tab.~\ref{tab:config}
we give the mass-weighted effective spin $\chi _\text{eff}$ that, in the equal-mass case, 
simply reduces to
\begin{equation}
\chi _\text{eff} = \frac{\chi^{A_\parallel} + \chi^{B_\parallel}}{2},
\end{equation}
with $\chi^{A_\parallel,B_\parallel}$ being the projection of the dimensionless spin vector
along the orbital angular momentum direction; 
and the effective spin-precession parameter $\chi _p$ that, in the equal-mass case, is defined as
\begin{equation}
\chi _p = \max (\chi ^{A_\perp},\chi ^{B_\perp}),
\end{equation}
where $\chi ^{A_\perp,B_\perp}$ is the magnitude of the component of the dimensionless spin
vectors perpendicular to the orbital angular momentum. 
Both spin measurements $\chi_{\rm eff},\chi_p$ are commonly used in GW data 
analysis~\cite{TheLIGOScientific:2017qsa,Abbott:2018wiz,LIGOScientific:2018mvr} for BNS systems 
and therefore seem to be a natural choice for a comparison with our simulations.

\begin{figure*}[htb]
\vspace{-1cm}
\begin{minipage}[t]{0.25\linewidth}
\includegraphics[width=0.98\linewidth]{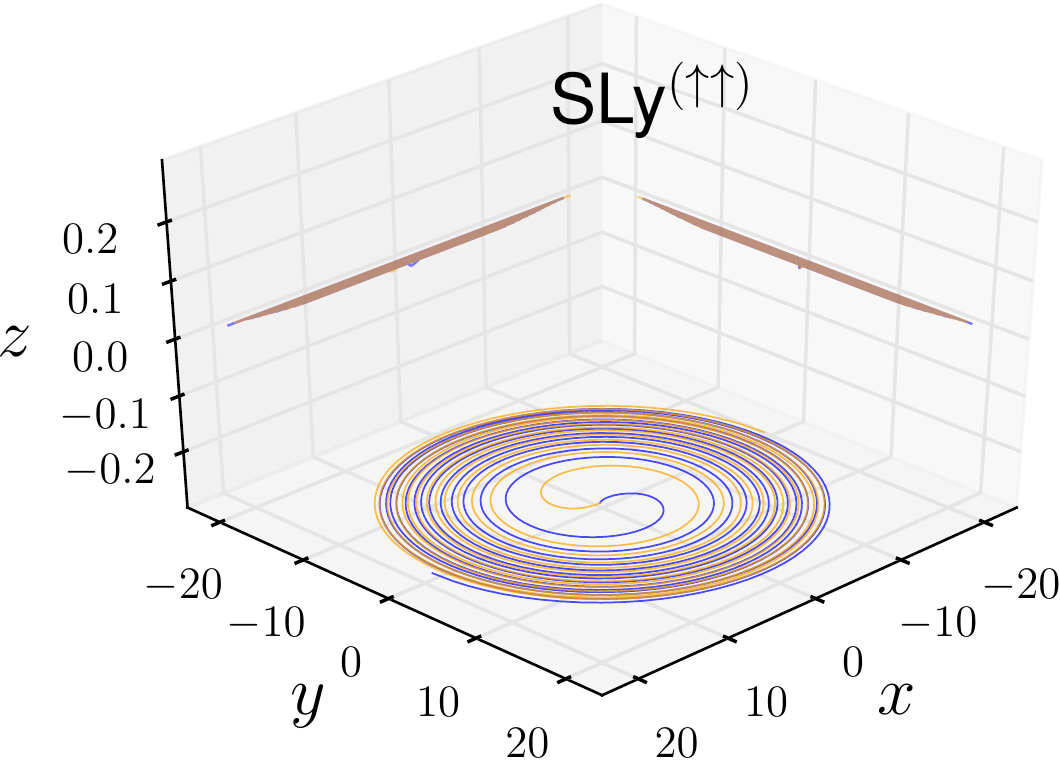}
{}
\end{minipage} \hspace{0.5cm} 
\begin{minipage}[t]{0.25\linewidth}
\includegraphics[width=\linewidth]{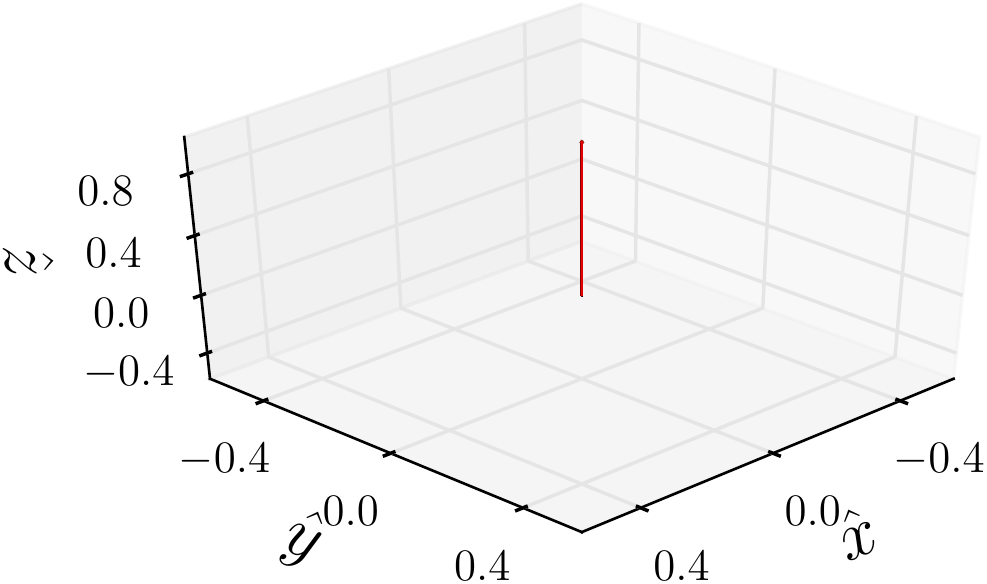}
\end{minipage} \hspace{0.5cm}
\begin{minipage}[t]{0.25\linewidth}
\includegraphics[width=1.05\linewidth]{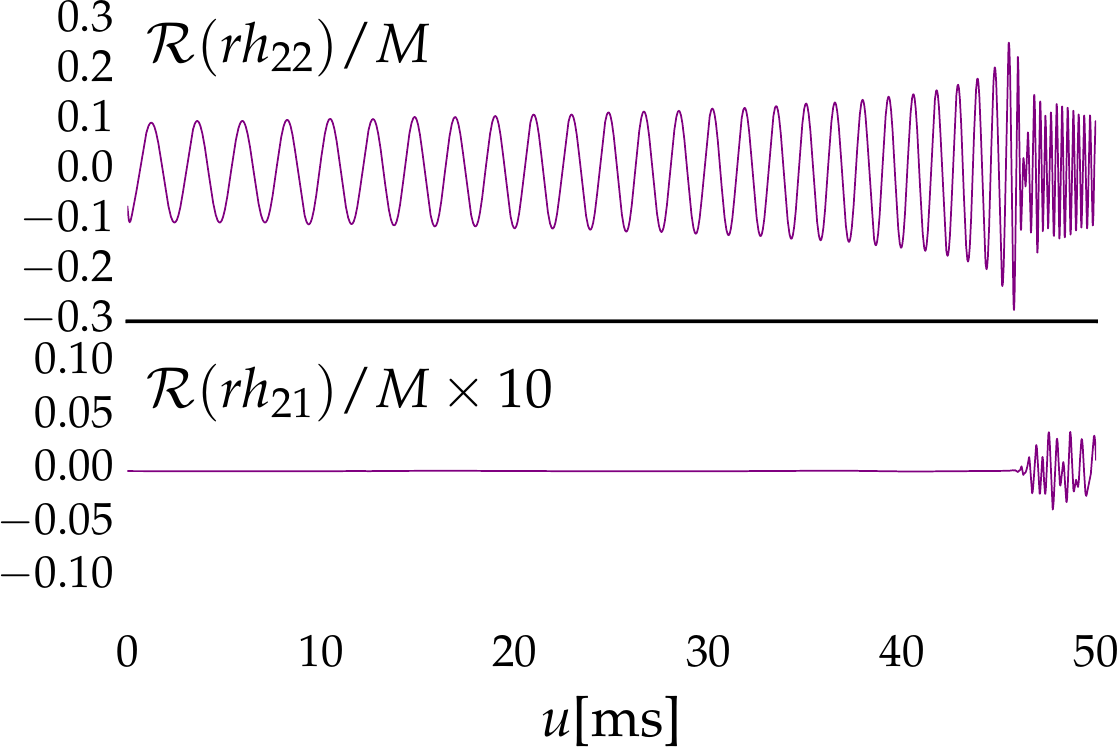}
\end{minipage} \hfill

\begin{minipage}[t]{0.25\linewidth}
\includegraphics[width=0.975\linewidth]{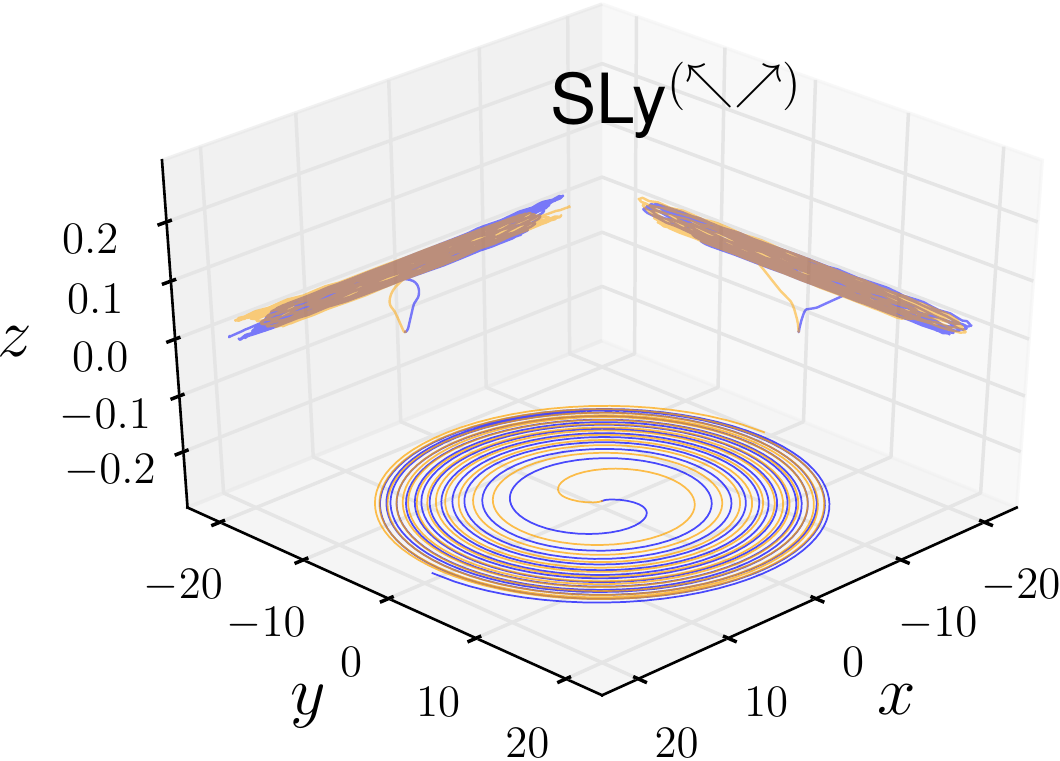}
{}
\end{minipage} \hspace{0.5cm}
\begin{minipage}[t]{0.25\linewidth}
\includegraphics[width=\linewidth]{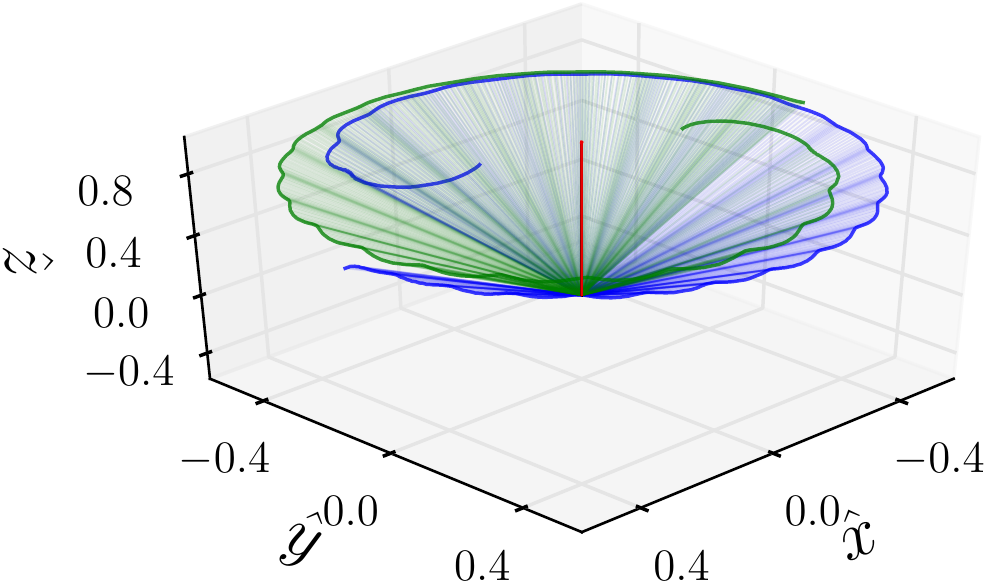}
\end{minipage} \hspace{0.5cm}
\begin{minipage}[t]{0.25\linewidth}
\includegraphics[width=1.05\linewidth]{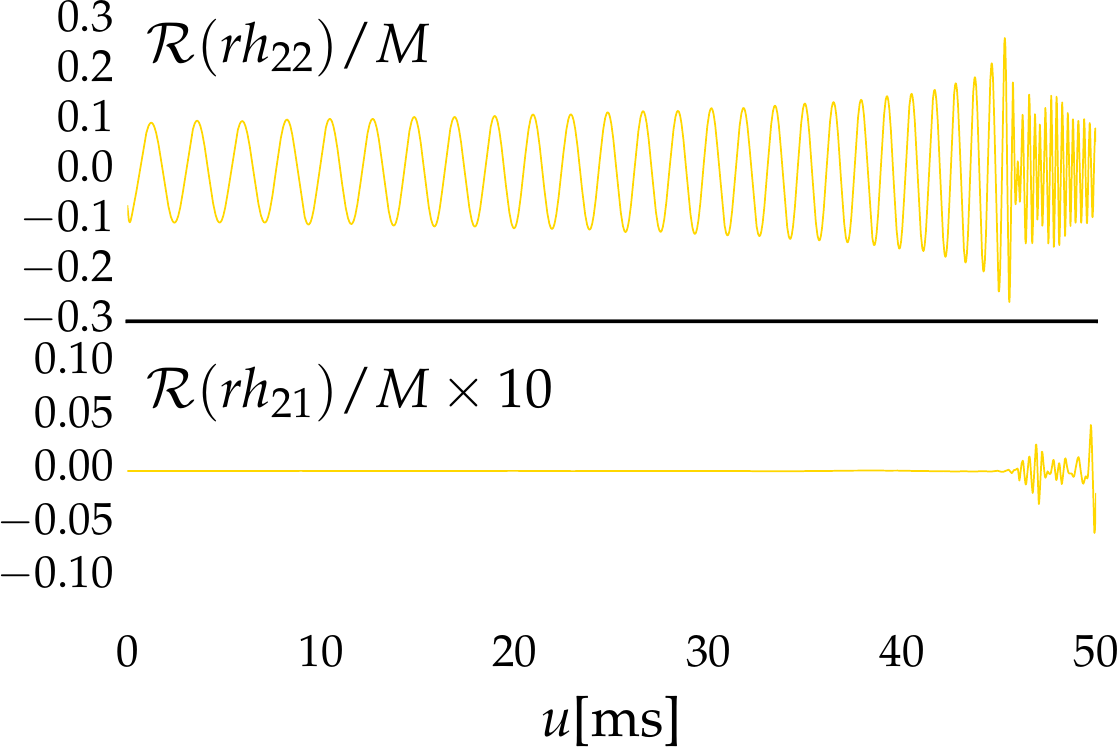}
\end{minipage} \hfill

\begin{minipage}[t]{0.25\linewidth}
\includegraphics[width=0.975\linewidth]{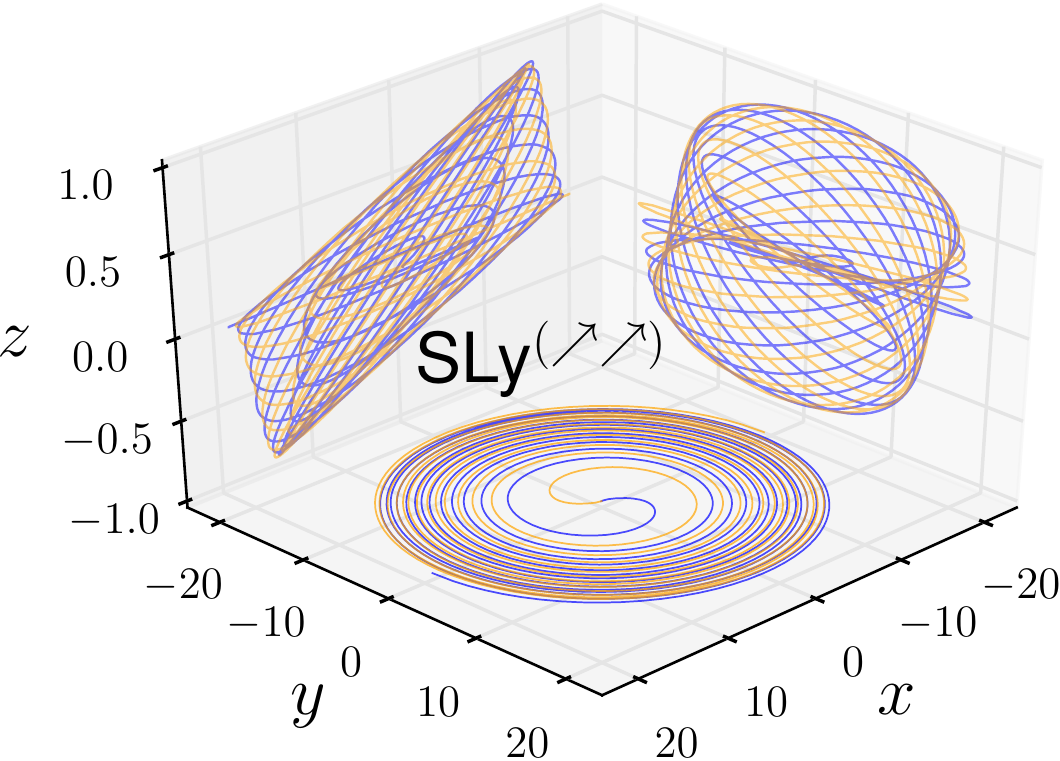}
{}
\end{minipage} \hspace{0.5cm}
\begin{minipage}[t]{0.25\linewidth}
\includegraphics[width=\linewidth]{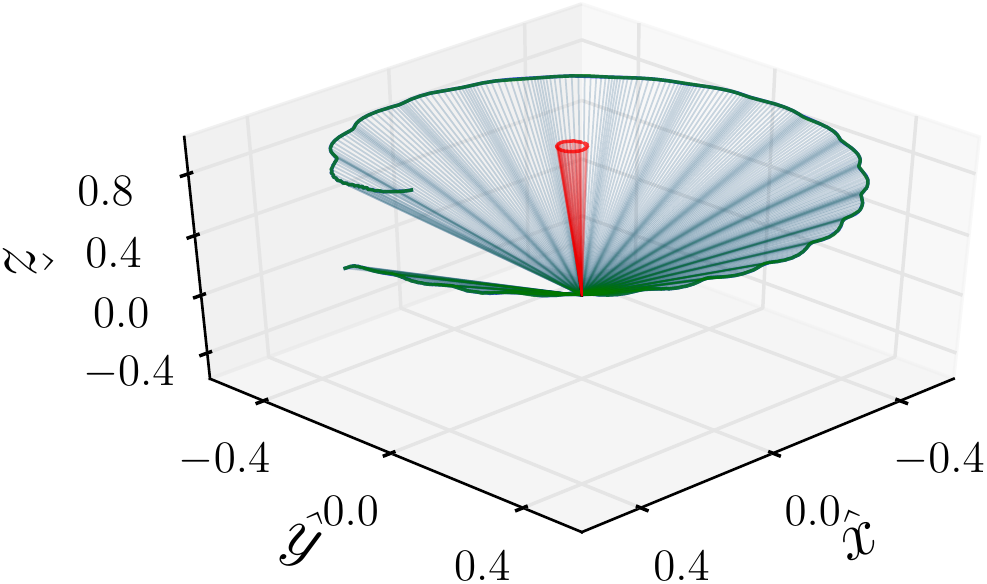}
\end{minipage} \hspace{0.5cm}
\begin{minipage}[t]{0.25\linewidth}
\includegraphics[width=1.05\linewidth]{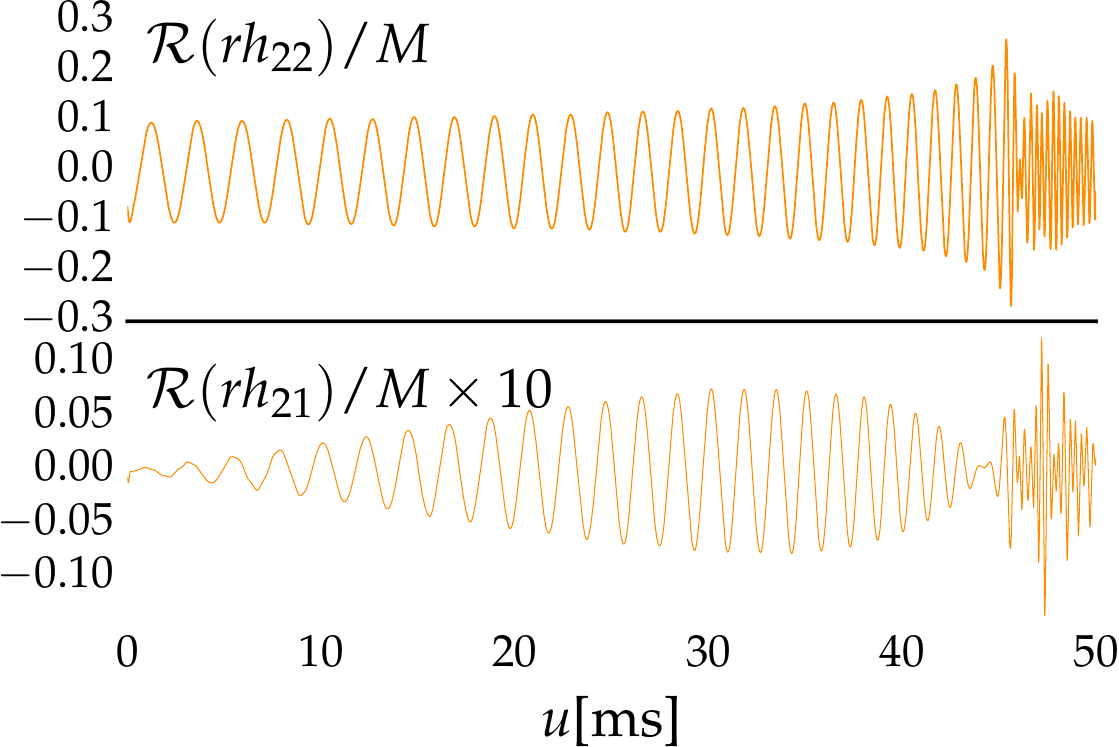}
\end{minipage} \hfill

\begin{minipage}[t]{0.25\linewidth}
\includegraphics[width=0.975\linewidth]{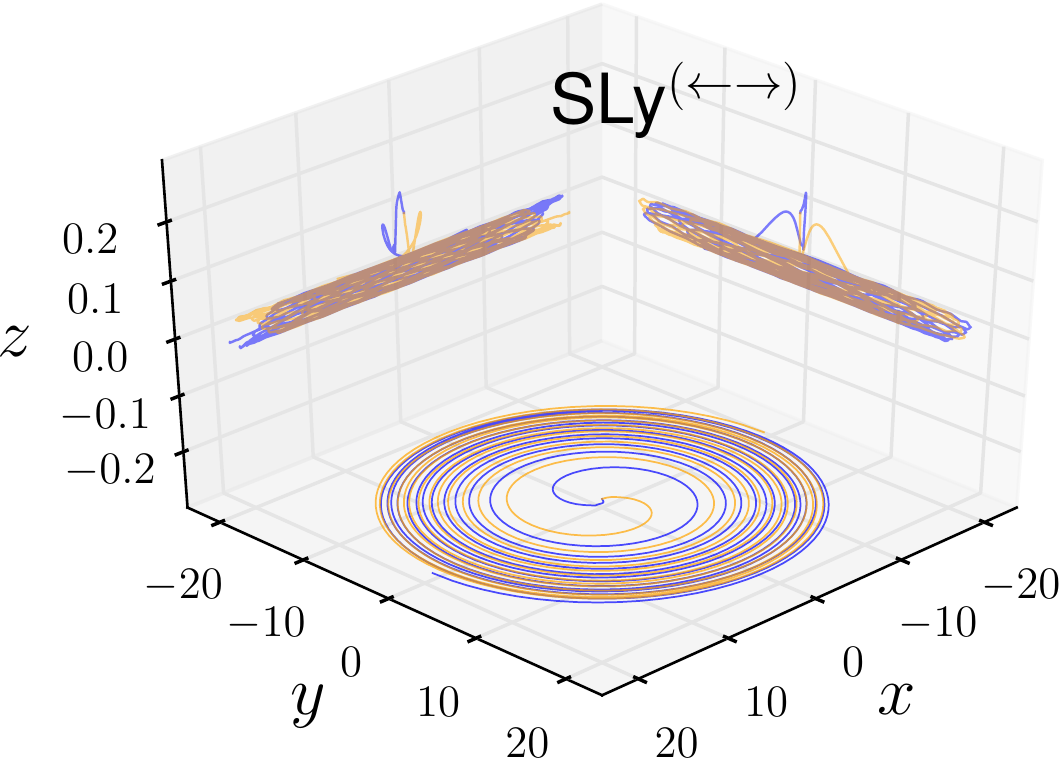}
{}
\end{minipage} \hspace{0.5cm}
\begin{minipage}[t]{0.25\linewidth}
\includegraphics[width=\linewidth]{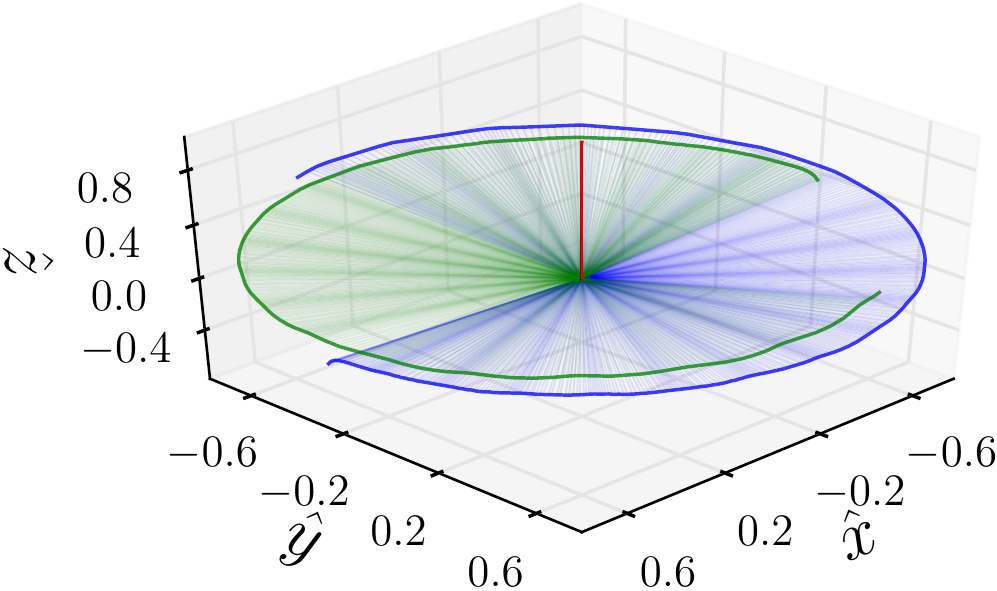}
\end{minipage} \hspace{0.5cm}
\begin{minipage}[t]{0.25\linewidth}
\includegraphics[width=1.05\linewidth]{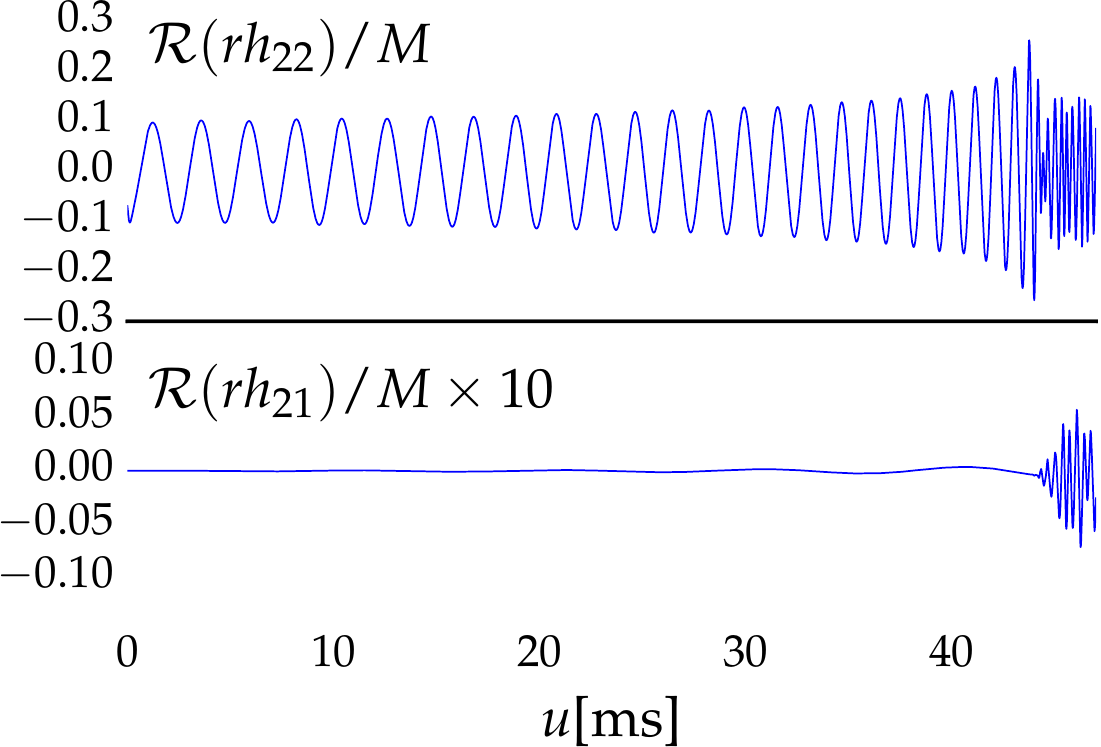}
\end{minipage} \hfill

\begin{minipage}[t]{0.25\linewidth}
\includegraphics[width=0.975\linewidth]{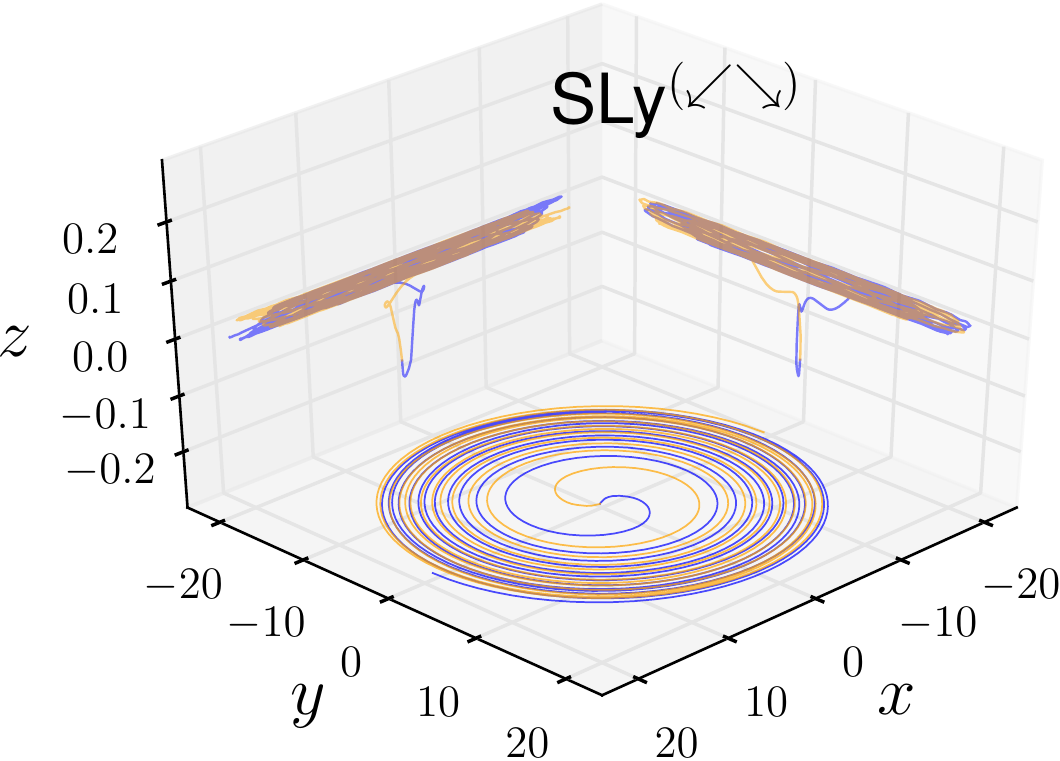}
{}
\end{minipage} \hspace{0.5cm}
\begin{minipage}[t]{0.25\linewidth}
\includegraphics[width=\linewidth]{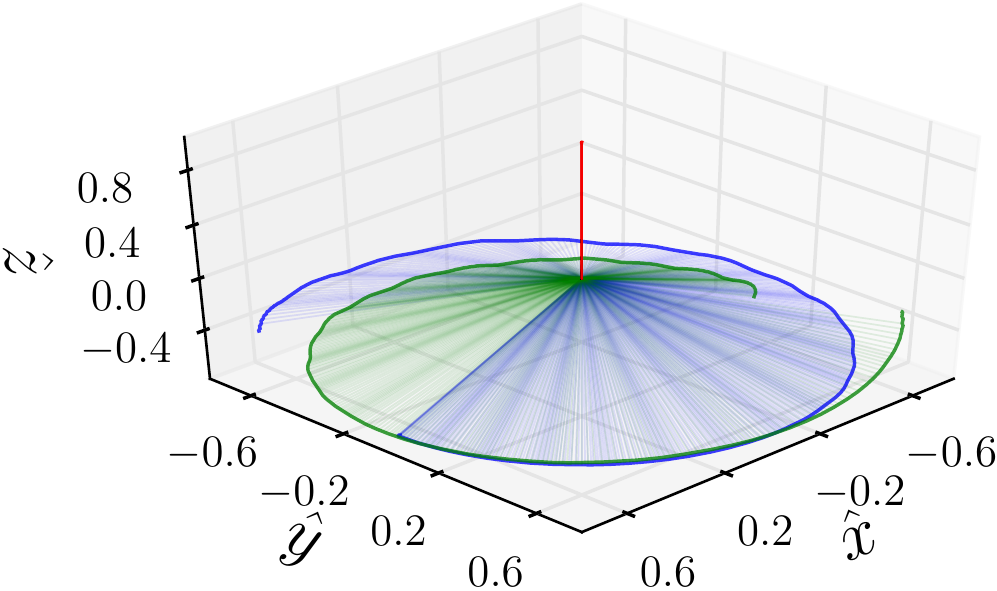}
\end{minipage} \hspace{0.5cm}
\begin{minipage}[t]{0.25\linewidth}
\includegraphics[width=1.05\linewidth]{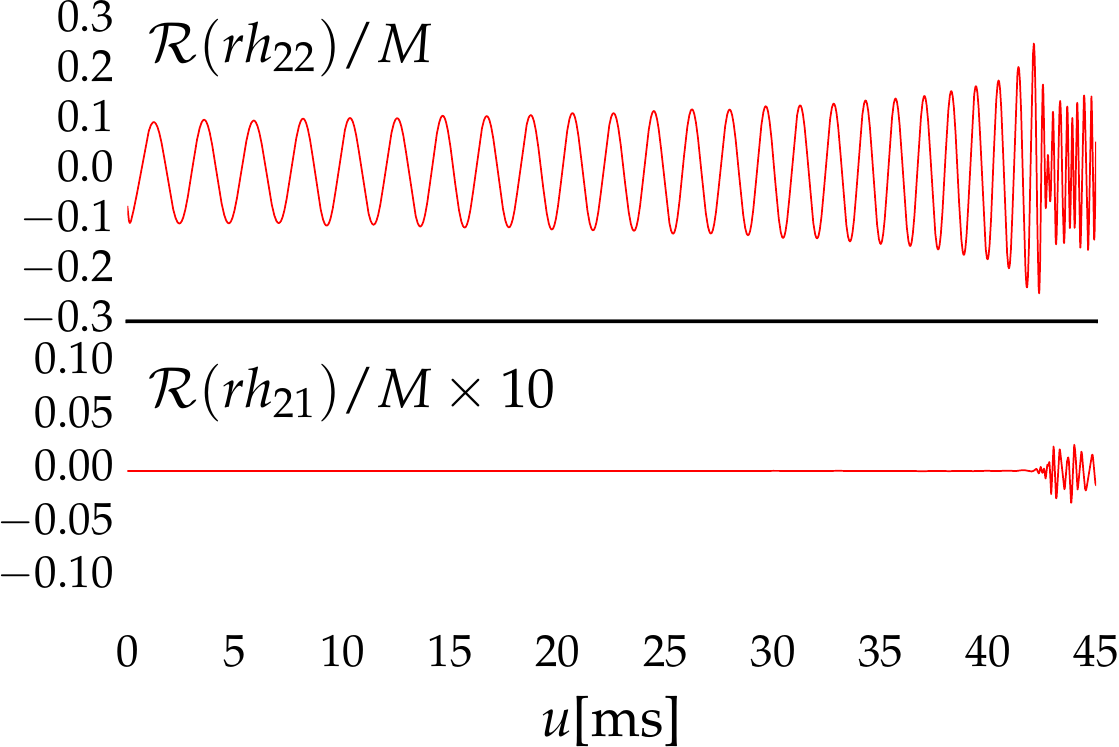}
\end{minipage} \hfill

\begin{minipage}[t]{0.25\linewidth}
\includegraphics[width=0.975\linewidth]{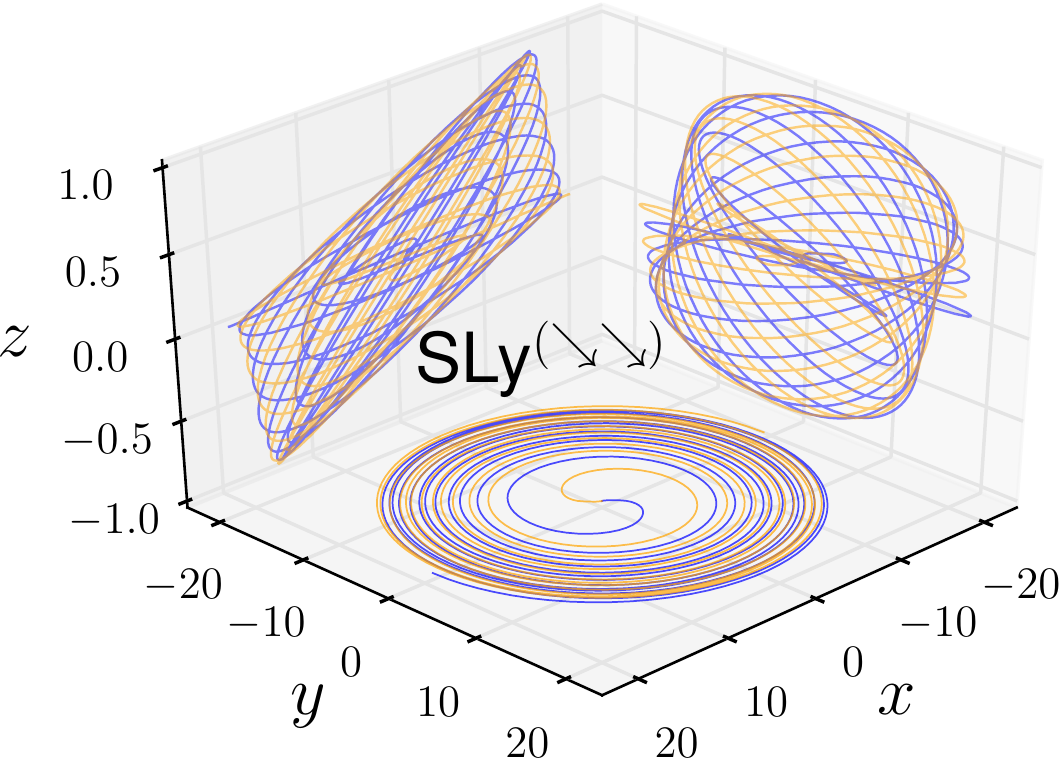}
{}
\end{minipage} \hspace{0.5cm}
\begin{minipage}[t]{0.25\linewidth}
\includegraphics[width=\linewidth]{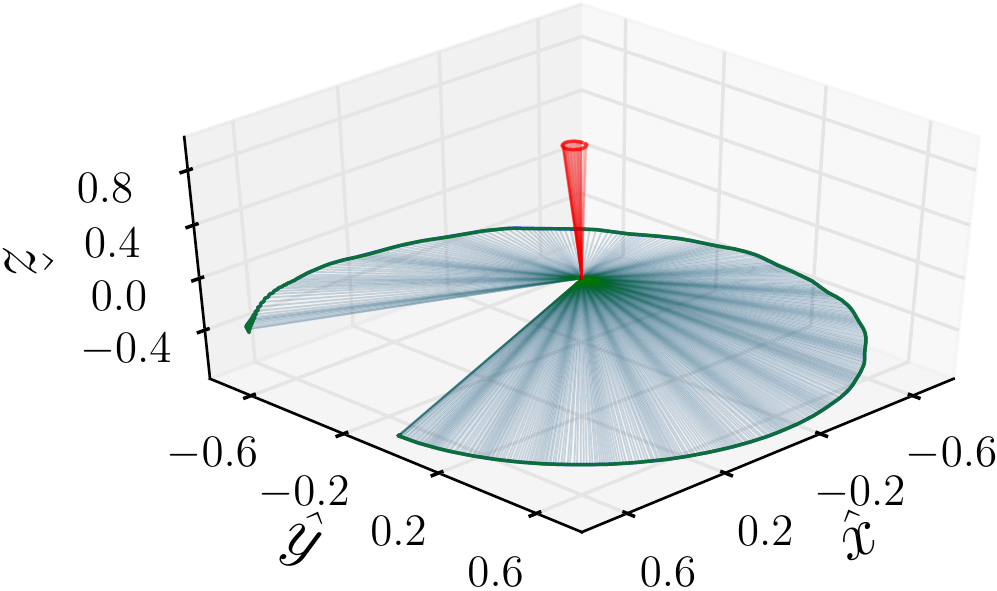}
\end{minipage} \hspace{0.5cm}
\begin{minipage}[t]{0.25\linewidth}
\includegraphics[width=1.05\linewidth]{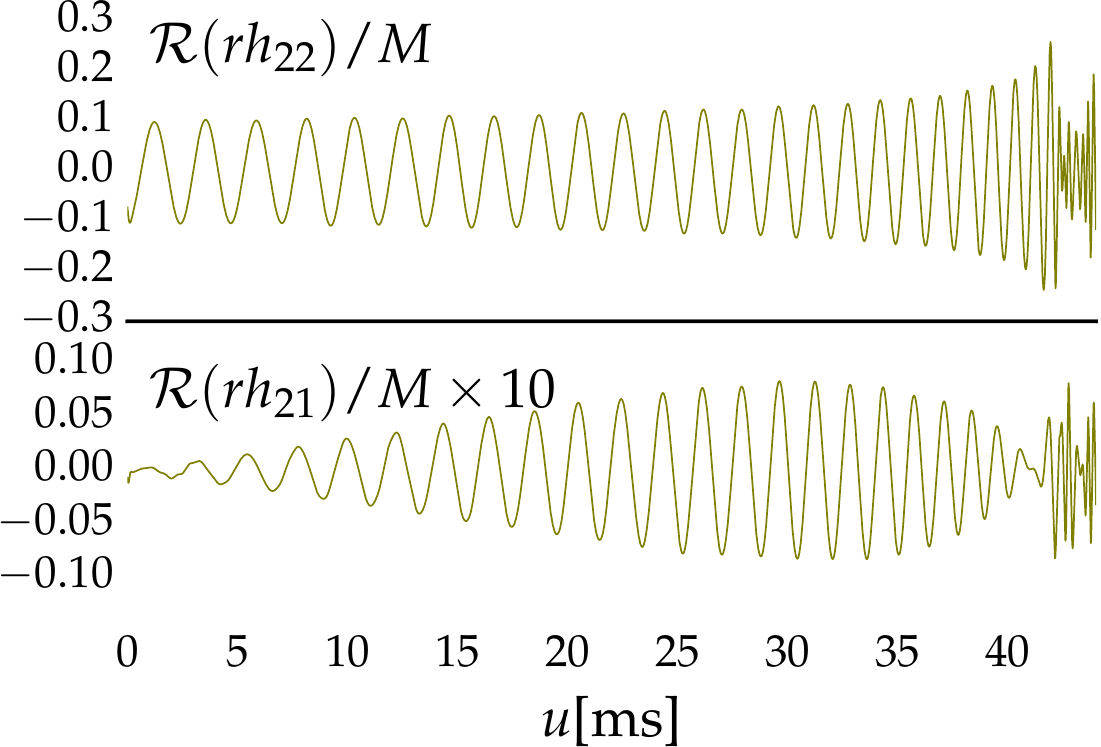}
\end{minipage} \hfill

\begin{minipage}[t]{0.25\linewidth}
\includegraphics[width=0.975\linewidth]{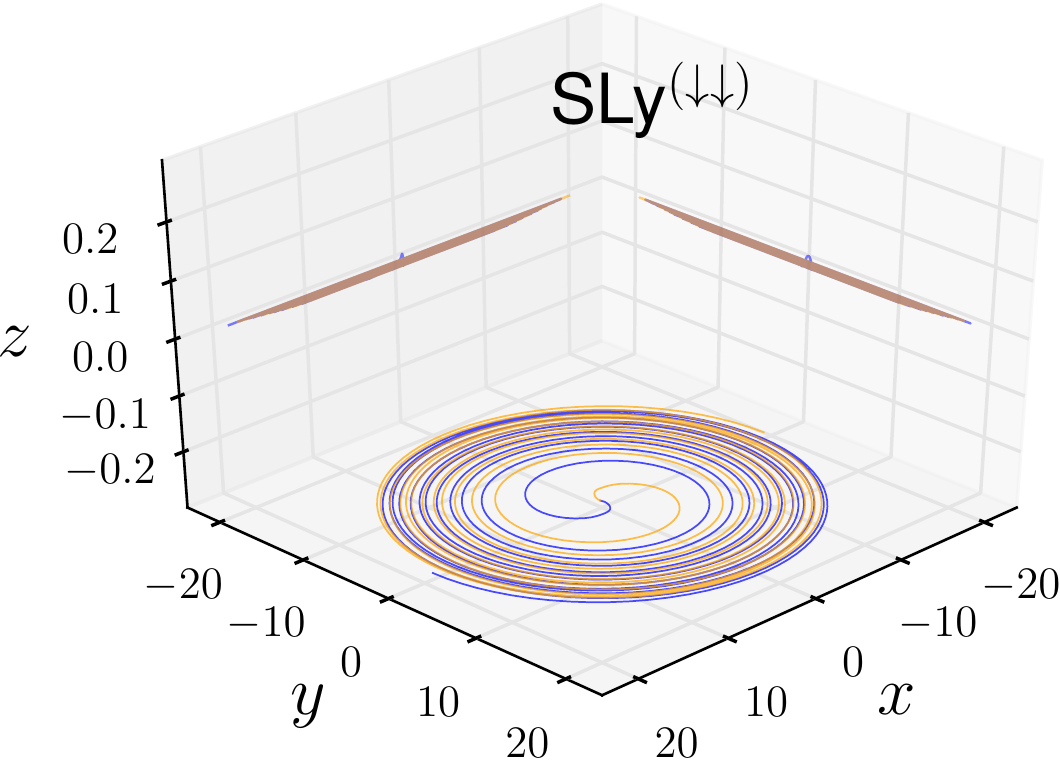}
{}
\end{minipage} \hspace{0.5cm}
\begin{minipage}[t]{0.25\linewidth}
\includegraphics[width=\linewidth]{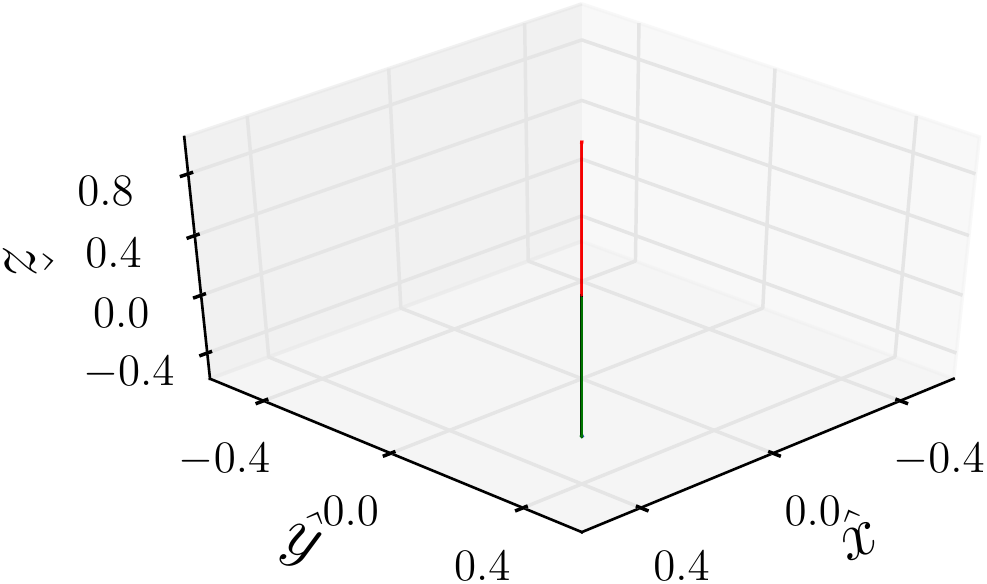}
\end{minipage} \hspace{0.5cm}
\begin{minipage}[t]{0.25\linewidth}
\includegraphics[width=1.05\linewidth]{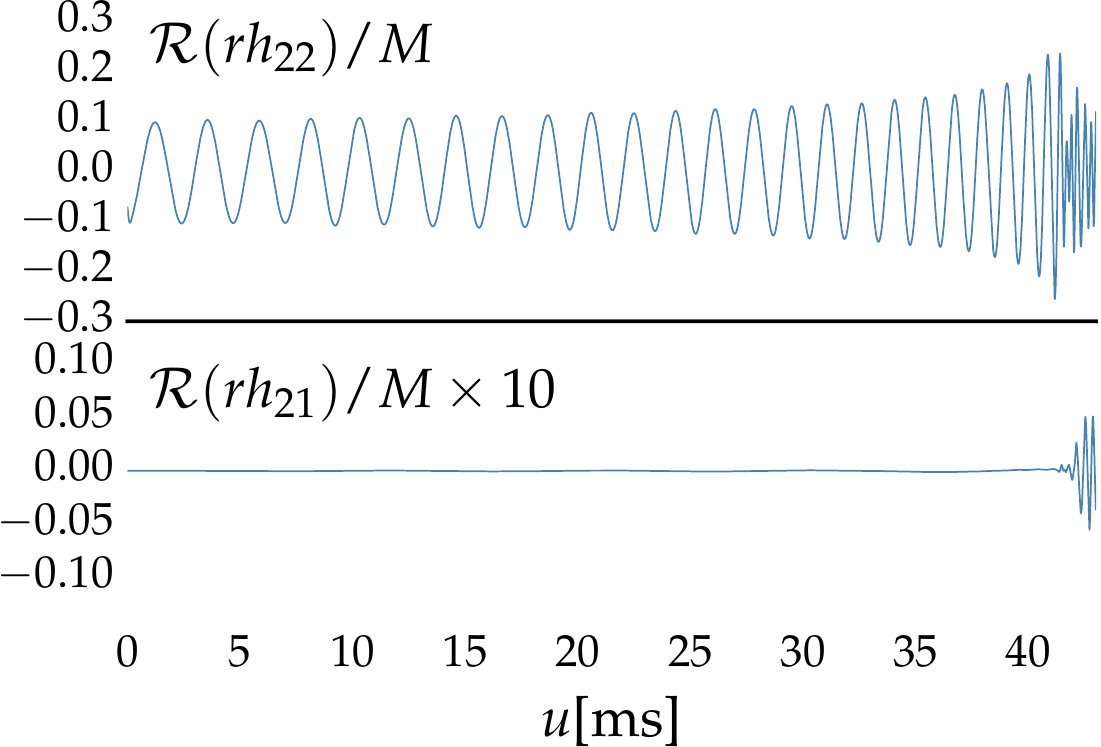}
\end{minipage} \hfill
\caption{Orbital dynamics and GW emission for all
simulations.
\emph{Column 1} shows the coordinate tracks of each NS
in the binary. \emph{Column 2} shows the corresponding precession cones.
The spin evolution of the individual stars (blue and green),
and the orbital angular momentum of the system (red) are
shown here. \emph{Column 3} shows the (2,2)- and
(2,1)-modes of the GW strain $rh$.}
\label{fig:dynafigs}
\end{figure*}

\section{Dynamics}
\label{sec:dyn}

\subsection{Qualitative discussion}

We start our investigation with a qualitative discussion about all considered systems. 
For this purpose, we present in Fig.~\ref{fig:dynafigs} the tracks of the stars (left panels), 
the precession cones of the individual spins and the orbital angular momentum (middle panels), 
and the (2,2)- and (2,1)-modes of the GW signal (right panels). 
The individual rows refer to the different configurations. 

Precession effects for~\Snee~and~\Ssee~are largest due to the misaligned initial spins, 
which leads to a clearly visible motion of the binaries along the $z-$axis. 
In addition, the precession cone of the orbital angular momentum has the largest opening angle which confirms our observation
that these systems undergo a precessing motion.
Moreover, a clear modulation in GW amplitude due to precession can be seen in the (2,1)-mode of the GW signal. 

Other configurations, such as~\Swe~have clearly different dynamics. 
Even though the initial spins for this simulation are misaligned, no 
characteristic precession effect is visible for the orbital angular momentum. 
As the spins lie in the orbital plane and are opposite and of equal magnitude, any $z-$motion of the stars is in the same direction. This results in a ``bobbing'' motion of the 
orbital plane (rather than the ``wobbling'' motion that is typical for precession). 
These findings are supported by the corresponding precession cone, which shows no precession of the 
orbital angular momentum. Furthermore, no precession effects are present in the (2,1)-mode of 
the GW signal due to the symmetry of this system. 
However, we find clearly that the individual spins are precessing; cf.~blue and green lines in the middle panels. 

Similar symmetry arguments can be used to explain why the other symmetrically misaligned 
simulations show ``bobbing'' motion in the $z$-direction, but no precession of the orbital plane 
like the~\Snee~and the~\Ssee~cases.

\begin{figure}[t]
\includegraphics[width=0.5\textwidth]{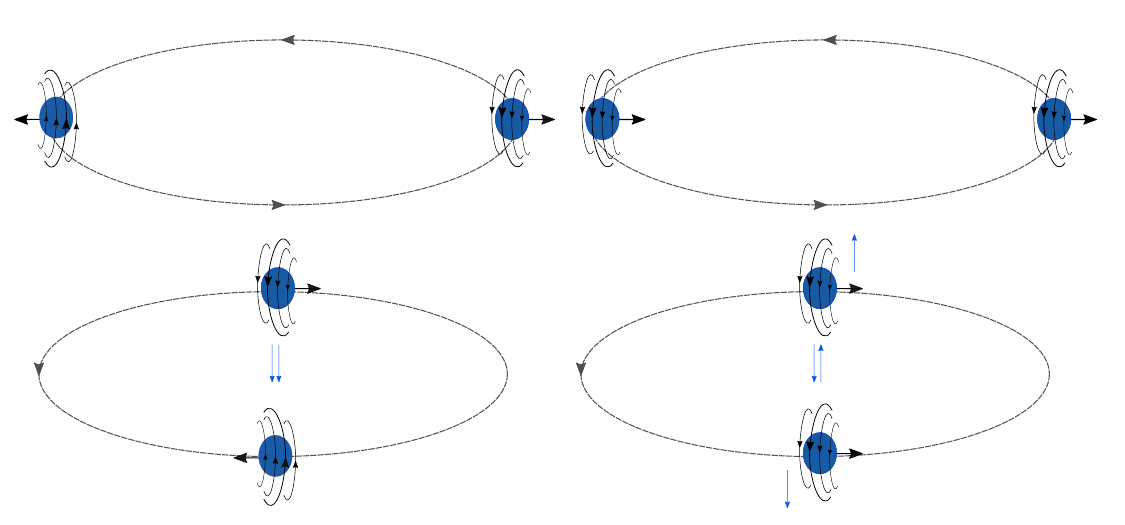}
\caption{A schematic of the \emph{Lense-Thirring effect} in a binary system.
\emph{Left column:} A system with symmetrically misaligned-spins of the NSs A and B
with respect to the orbital angular momentum direction.
\emph{Right column:} A system with asymmetrically misaligned-spins of the NSs A and B.}
\label{fig:frame_drag}
\end{figure}

Interestingly, the motion
of the orbital plane, ``wobbling'' or ``bobbing''
for the spin misaligned systems
can be explained by considering the general relativistic
frame-dragging effect or specifically the
Lense-Thirring (LT) effect~\citep{Morsink:1999}.
Due to this effect, a rotating mass in general relativity
influences the motion of objects in its vicinity i.e.,
the rotating mass ``drags along" spacetime in its vicinity.
In Fig.~\ref{fig:frame_drag} we show a schematic of the
frame dragging due to the NS spins.
\emph{Top row panels} show the initial configurations for setups
with symmetrically misaligned-spin (\emph{left column}) and with
asymmetrically misaligned-spin (\emph{right column}) for the NSs
A and B. The blue circles represent the two NSs, the black arrows show
their spin directions and the dragging of the spacetime is depicted as
the circular rings around the NSs. In the \emph{top
row panel} scenario, the stars will feel no LT effect i.e., the dragging
due to each other's spin rotations. The spin orientation of the stars
changes very slowly, so that a
quarter of an orbit later they will still be pointed in essentially the same
direction. This scenario is depicted in the \emph{bottom row panels}.
This time for the symmetrically misaligned system,
star B will feel the LT effect due to A in the direction that points into
the orbital plane. Since the spin of star B is pointed in the opposite direction,
the LT effect on star A will be in the same direction as on star B i.e., into the
orbital plane as shown in the \emph{bottom left panel}. The net result will now push
the entire orbital plane in this direction, which is perpendicular to the orbital plane.
Half an orbit later, the effect will be in the opposite direction; the resulting
motion is an oscillation of the orbital plane in the perpendicular direction
i.e., a ``bobbing'' motion. 
Similarly, for the asymmetrically misaligned-spin system (\emph{right column});
stars A and B will be pushed in directions opposite to one another. This results
in a zero net force on the orbital plane as shown in the \emph{bottom right panel} and
a nonzero torque that tilts the orbital plane, which over time causes the ``wobbling''
motion. Therefore, the misaligned-spins of the NSs either produces a
torque or a net force on the orbital plane giving rise
to either ``wobbling'' or ``bobbing'' motions respectively.
Additionally, two important observations can be made based on Fig.~\ref{fig:dynafigs}.

\begin{figure}[t]
\includegraphics[width=0.48\textwidth]{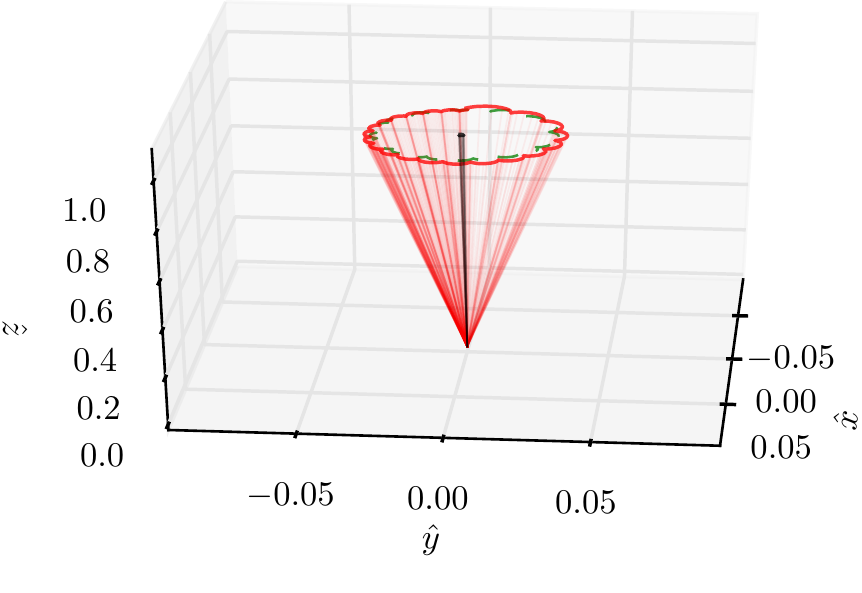}
\caption{Precession cone for the~\Snee~configuration
for the orbital angular momentum
(red). Additionally, as green dashed line we show also the precession cones for
$(\hat{L}_x,\hat{L}_y,\hat{L}_z)$ for the~\Ssee~configuration.
The opening angles for both the configurations are almost identical,
due to the symmetry of the systems.}
\label{fig:precession_Orb}
\end{figure}

First, for the~\Suu~and~\Sdd~configurations there is no precession as their initial spins are (anti-) aligned with the orbital angular momentum. Moreover, the orbital hang-up (speed-up) effect~\citep{Campanelli:2006uy,Bernuzzi:2013rza},
i.e., the fact that spin-aligned systems merge later and vice versa, is clearly visible in the GW signal with respect to the peak time in the amplitude at the merger. This effect also holds for the misaligned systems that have an effective (anti-) aligned spin components with respect to the orbital angular momentum of the system. The exact merger times can be found in Tab.~\ref{tab:remnant} for the R3 setups.
Second, apart from precession, the spin misaligned
systems also show nutation, i.e., small
oscillations in the precession cones for
the individual spins (blue and green) as seen in column 2 of~Fig.~\ref{fig:dynafigs}.
The nutation happens on a much
shorter timescale than the precession motion.
These nutation cycles are clearly visible
for the individual spins for the~\Snw~and~\Snee~cases
but are also present for the~\Ssw~and~\Ssee~cases.
We also show a comparison of the precession cones of the orbital
angular momentum for \Snee~and \Ssee~in Fig.~\ref{fig:precession_Orb}.

\subsection{Energetics}

\begin{figure}
\includegraphics[width=0.5\textwidth]{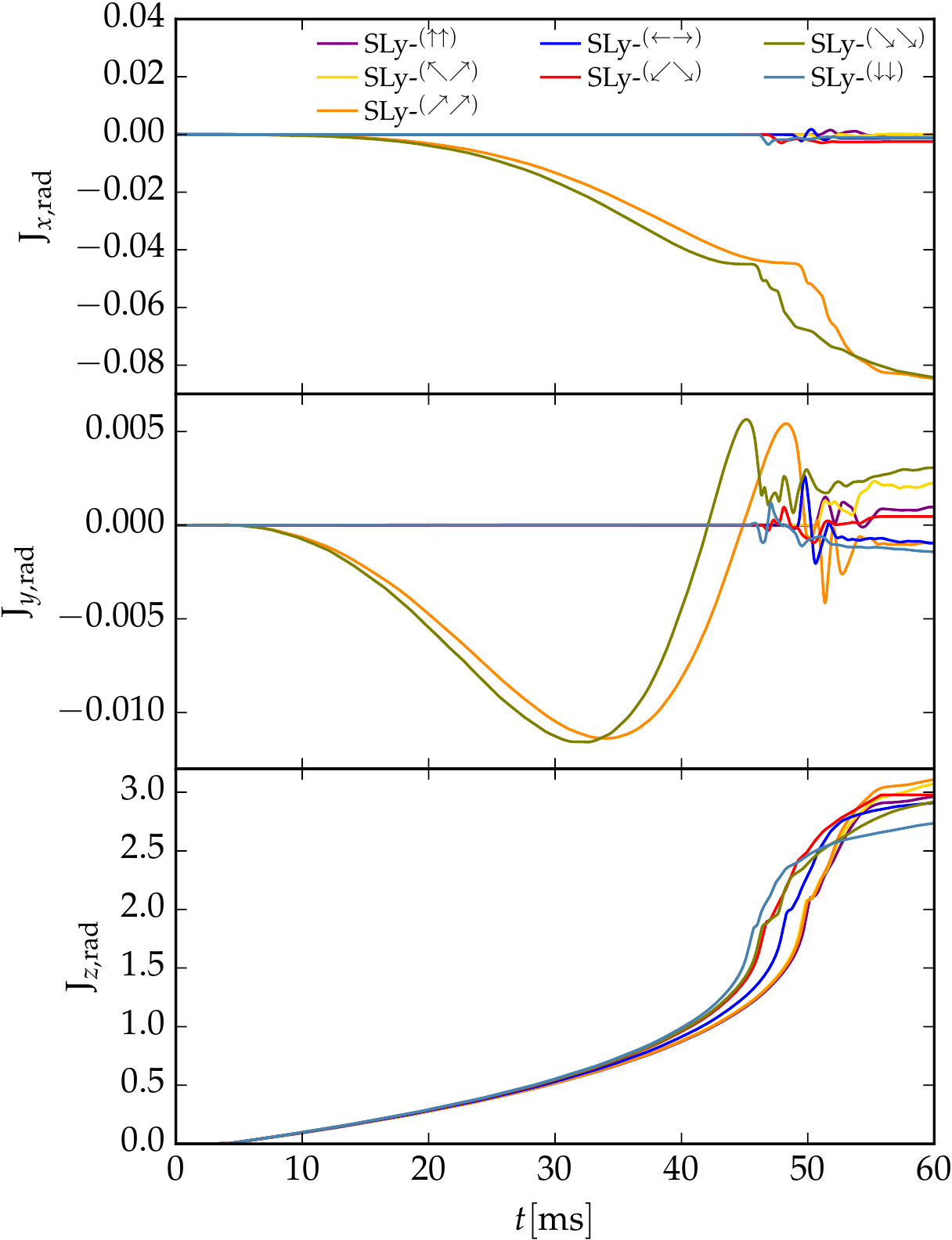}
\caption{Radiated angular momentum ($\vec{\bf{J}}_{\rm rad}$) in GWs computed
using the relations given in Appendix~\ref{app:EJ_formulas}.
For symmetrically misaligned configurations the angular momentum
is only radiated in the $z-$component and the
$x,y-$components remain identically zero
during the inspiral. Whereas for the
asymmetrically misaligned systems there is
radiation in all the components.}
\label{fig:angular_mom_rad}
\end{figure}

We study the conservative dynamics for all the
configurations presented in this article by
computing the reduced binding energy,
\begin{equation}
E_{b}=\frac{M_{\rm ADM}(t_0)-E_{\rm rad}-M}{\nu \ M}, 
\end{equation}
and the specific orbital angular momentum,
\begin{equation}
\ell =\frac{|{\vec{\bf{J}}_{\rm ADM}}(t_0)- {\vec{\bf{S}}_A}(t_0) -
{\vec{\bf{S}}_B}(t_0) -\vec{\bf{J}}_{\rm rad}|}{\nu\ M^2}. 
\end{equation}
Here $\nu:= M^A M^B/M^2$ is the symmetric mass ratio, 
$E_{\rm rad},\vec{\bf{J}}_{\rm rad}$ are the emitted energy and angular momentum in the radiated GWs,
and $M_{\rm ADM}, \vec{\bf{J}}_{\rm ADM}$ denote the 
ADM mass and angular momentum at the beginning of the simulation (i.e.~at $t = t_0$),
${\vec{\bf{S}}_A}(t_0)$ and ${\vec{\bf{S}}_B}(t_0)$ are estimated from the initial data (Tab.~\ref{tab:config}),
and ${\vec{\bf{S}}}_{A,B} = (M^{A,B})^2 \chi^{A,B} \hat{\chi}^{A,B}$.
In Appendix~\ref{app:EJ_formulas}, we present a few details
about the postprocessing step
for the computation of the radiated energy and
angular momentum in GWs extracted in numerical relativity
simulations employing the \texttt{BAM} code. In Fig.~\ref{fig:angular_mom_rad},
we show the computed angular momentum for all the simulated configurations.
One finds that for the symmetrically misaligned
configurations (\Swe, \Snw, \Ssw) the angular momentum is
radiated only in the $z-$component whereas the
$x,y-$components remain identically zero
during the inspiral. For the
asymmetrically misaligned systems (\Snee and \Ssee) there is
radiation in the other components as well.

\begin{figure}
\includegraphics[width=0.5\textwidth]{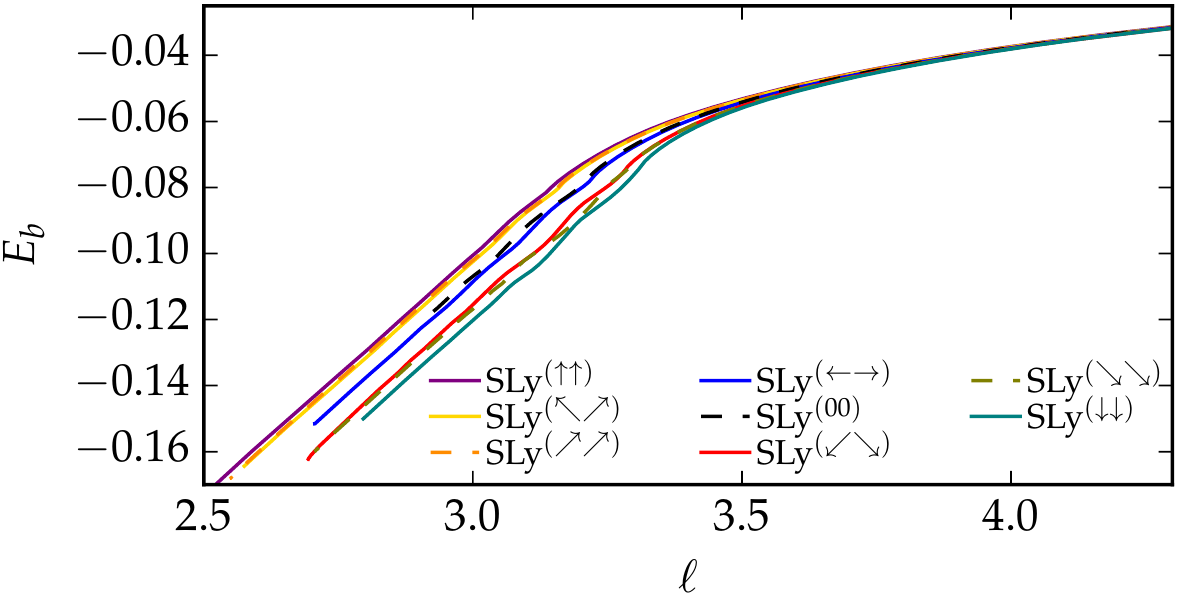}
\caption{Reduced binding energy $E _b$ as a function
of the specific orbital angular momentum $\ell$
for all configurations considered in the
article. Additionally, we also include the curve
for an irrotational case~\Soo~taken
from the CoRe Database (ID:-\texttt{BAM}:0095:R02) for comparison.
As expected, the irrotational curve matches nicely to
the~\Swe~case.}
\label{fig:Eell_curve}
\end{figure}

In Fig.~\ref{fig:Eell_curve} we show the $E$-$\ell$ curve
for all the configurations for the highest resolution (R4),
cf.~Tab.~\ref{tab:grid}.
For comparison we also show the curve
for an irrotational configuration
\Soo\ (`black line') with the same masses and EOS.
The irrotational setup corresponds to ``\texttt{BAM}:0095:R02"
from the CoRe database~\cite{CoRe,Dietrich:2018phi}. For the
early inspiral part of the dynamics
(large $E$ and $\ell$), we find that
the $E$-$\ell$ curves are very similar for all setups, 
which is caused by the fact that the main contribution, 
the point-mass contribution, is identical for all systems. 

\begin{figure}
\includegraphics[width=0.5\textwidth]{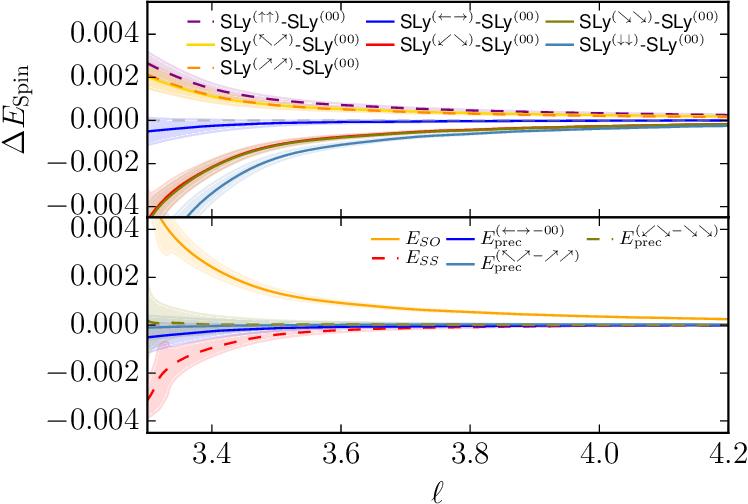}
\caption{Top panel: Estimate of the spin
orientation effects on the
conservative dynamics by taking
the difference between all the
configurations and the \Soo
(irrotational case taken from CoRe Database)
configuration. The shaded
region marks the difference in results
obtained with a lower
resolution and takes into account the
uncertainty of the initial
data. Bottom panel: Spin and orbital
contributions to the
binding energy estimated following the
discussion in the text.}
\label{fig:Ej_contrib}
\end{figure}

During the late inspiral part, when the stars
come close to each other, due to the emission of
energy and angular momentum, a clear difference is present as seen
in~Fig.~\ref{fig:Eell_curve}. 
Throughout the simulation, 
the $E$-$\ell$ curve for the irrotational configuration~\Soo~and the effectively 
zero-spin configuration ($\chi_{\rm eff} =0$)~\Swe~clearly demarcate the
effectively aligned spin and the effectively
antialigned spin configurations. 
In general, aligned spin configurations are less bound
while the antialigned spin configurations are more bound
than the corresponding irrotational setup, cf.~Fig.~\ref{fig:Ej_contrib}
top panel. 

To better disentangle the different contributions
to the total binding energy due to the 
spin~\cite{Bernuzzi:2013rza,Dietrich:2016lyp},
we assume that it consists of a
nonspinning contribution including tidal
effects $E _0$, a spin-orbit $E_{\text{SO}}$ contribution,
and a spin-spin contribution $E _{\text{SS}}$,
\begin{equation}
E_b = E_0 + E_{\text{SO}} + E_{\text{SS}} + \mathcal{O}(S^3).
\label{eq:Eb_contrib}
\end{equation}
In general, the spin-orbit ($\text{SO}$) interaction is at leading order
$\propto \vec{\bf L}\cdot \vec{\bf S}_i/r^3$, see~\citep{Kidder:1992fr}.
The $\text{SO}-$interaction term is either repulsive or
attractive, i.e., positive or negative, according to the sign of
$\sum_{i=1}^2\vec{\bf L}\cdot \vec{\bf S}_i$.
The spin-spin term includes the self-spin term
(of the form $\vec{\bf S}_i \cdot \vec{\bf S}_i$) and an interaction term
(of the form $\vec{\bf S}_i \cdot \vec{\bf S}_j\ (i \neq j)$)
between the two spins. The spin-spin interaction term in particular
is $ \propto [3(\vec{\bf n}\cdot {\vec{\bf S}}_1)(\vec{\bf n}\cdot{\vec{\bf S}}_2)-
({\vec{\bf S}}_1\cdot{\vec{\bf S}}_2)]/r^3$ (with $\vec{\mathbf{n}}$ denoting the
unit vector pointing from one star to the other
and $r$ being the distance between the stars),
see e.g.~\citep{Kidder:1992fr}.
Note that the first term in the interaction term
is zero for the (anti-) aligned configurations
and the remaining term $ \propto -({\vec{\bf S}}_1\cdot{\vec{\bf S}}_2)$
does not change sign if both spins flip.
We compute the spin-orbit term $E_{\text{SO}}$ as
\begin{equation}
 E_{\text{SO}} = \frac{E_b[{\rm SLy}^{(\uparrow \uparrow)}] - E_b[{\rm SLy}^{(\downarrow \downarrow)}]}{2},
\label{eq:E_SO}
\end{equation}
and estimate the complete spin-spin term, i.e., including
the interaction term and the self-spin term as,
\begin{equation}
 E_{\text{SS}} = \frac{E_b[{\rm SLy}^{(\uparrow \uparrow)}] + E_b[{\rm SLy}^{(\downarrow \downarrow)}]}{2}- 
 E_b[{\rm SLy}^{(0 0)}].
\label{eq:E_SS}
\end{equation}
The bottom panel in Fig.~\ref{fig:Ej_contrib} shows
these contributions to the binding energy.
We find that compared to the $\text{SO}$-interaction, the spin-spin term is almost
negligible during most of the inspiral and mostly within the uncertainty
of our data~\footnote{The error estimate in Fig.~\ref{fig:Ej_contrib}
is shown as shaded regions. It is obtained by
taking into account the finite resolution of
the simulations and is estimated from the difference
between R3 and R4 resolutions. For the irrotational case
we do not have exactly the same resolution data,
namely R3 and R4 used in this article but higher resolutions
(finest resolution boxes have $h=0.078 M_{\odot}$ and $h=0.118 M_{\odot}$).
Furthermore, an additional uncertainty of $10^{-5}$ is added
for accounting the errors coming in from the
initial data solver~\cite{Dietrich:2015pxa}.
The error bounds shown are obtained from error propagation
assuming errors from different configurations are uncorrelated.}.
The $\text{SO}$-contribution is the dominant contribution to the binding energy in our comparison, 
while in the very late inspiral tidal effects can dominate~\citep{Bernuzzi:2013rza}.
Intuitively, this is understandable based on the differences in the PN order of
the SO (1.5PN), spin-spin (2PN), and tidal effects (5PN).

To get a better understanding of potential precession effects, 
we also compute 
\begin{align}
E_{\rm prec}^{(\nwarrow \nearrow - \nearrow \nearrow )}
& = E_b[{\rm SLy}^{(\nwarrow \nearrow)}] - E_b[{\rm SLy}^{(\nearrow \nearrow )}],\\
E_{\rm prec}^{(\swarrow \searrow - \searrow \searrow )}
& = E_b[{\rm SLy}^{(\swarrow \searrow)}] - E_b[{\rm SLy}^{(\searrow \searrow )}],\\
E_{\rm prec}^{(\leftarrow \rightarrow - 00)}
& = E_b[{\rm SLy}^{(\leftarrow \rightarrow)}] - E_b[{\rm SLy}^{(00)}],
\end{align}
and show the results in the bottom panel of Fig.~\ref{fig:Ej_contrib}.
We find that the configurations (\Snw~\&~\Snee)
and (\Ssw~\&~\Ssee) are almost identical
with respect to their binding energy contribution.
Also the difference between the irrotational case and \Swe 
is not clearly resolved in our simulations. 
The reasons for this could be due to (i) the fact that the spins are rather small 
to show any distinguishable effect and (ii) that even the
highest resolution employed in the simulations presented
in this article falls short in resolving the differences
between those configurations. 
Therefore, even though the tracks and the precession cones, 
cf.~Fig.~\ref{fig:dynafigs}, show clear imprints of precession, 
the energetics does not shed light on the differences,
at least among the abovementioned pairs.

\subsection{Merger remnant}

\begin{figure}[t]
\includegraphics[width=0.5\textwidth]{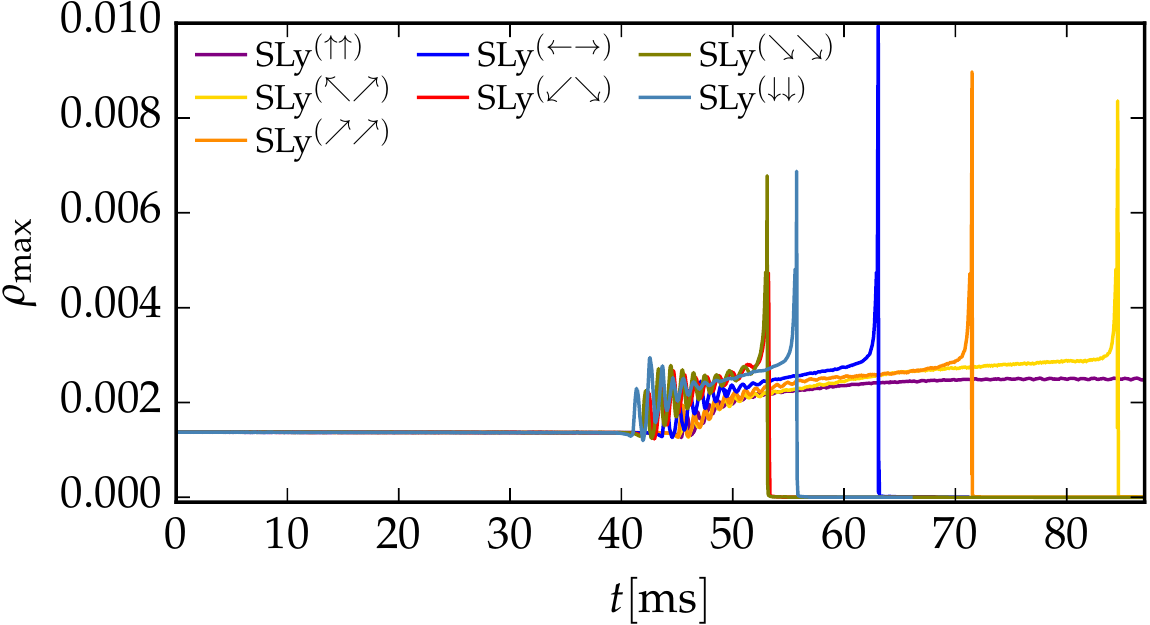}
\caption{Maximum of $\rho$ vs. coordinate time $t$. The cases that form a
black hole (BH) after the merger show a sharp change in the density where the
density drops to zero for such cases, because matter is removed inside
the BH.
Note that we report the merger remnant
properties for the R3 resolution setups as the simulations
could be evolved for longer times owing to the reduced computational costs.}
\label{fig:grhd_rho_max}
\end{figure}

\begin{table}[tp]
\caption{Properties of the merger remnant. 
The columns represent:
(i) the name of the configuration
(ii) the merger time in $M_{\odot}$ and in {\rm ms}
(iii) the lifetime, $\tau$, of the HMNS formed during our simulation,
given in $M_{\odot}$ and in {\rm ms};
(iv) the final mass of the BH, $M_\text{BH}$, if the HMNS collapsed during
our simulation;
the dimensionless spin of the final BH, $\chi_\text{BH}$
and the mass of the disk surrounding the BH,
$M_\text{disk}$. The different physical quantities are computed
for resolution R3.
Note that the case~\Suu~did not undergo collapse to a
BH during our simulation time and therefore the
corresponding quantities are marked as ``$-$''.}
\label{tab:remnant} 
\centering
\begin{small}
\begin{tabular}{c|cc|cc|ccc}
\toprule
Name & \multicolumn{2}{c|}{$t_{\rm merge}$} & \multicolumn{2}{c|}{$\tau$} & $M_\text{BH}$ & $\chi_\text{BH}$ & $M_\text{disk}$ \\
     & \begin{footnotesize} $[M_{\odot}]$  \end{footnotesize} & \begin{footnotesize} $[\rm ms]$  \end{footnotesize} & \begin{footnotesize} $[M_{\odot}]$  \end{footnotesize}& \begin{footnotesize} $[\rm ms]$  \end{footnotesize}& \begin{footnotesize} $[M_{\odot}]$  \end{footnotesize}&                  & \begin{footnotesize} $[M_{\odot}]$ \end{footnotesize}\\
\hline
\Suu & 9981 & 49.16 & $>12532$ & $>61.72$ & $-$ & $-$ & $-$ \\
\hline
\Snw & 9923 & 48.88 & 7074 & 34.84 & 2.37 & 0.57 & 0.215 \\
\hline
\Snee & 9919 & 48.86 & 4421 & 21.78 & 2.42 & 0.62 & 0.165 \\
\hline
\Swe & 9622 & 47.39 & 3062 & 15.08 & 2.40 & 0.59 & 0.167 \\
\hline
\Ssw & 9275 & 45.68 & 1394 & 6.87 & 2.45 & 0.62 & 0.113 \\
\hline
\Ssee & 9230 & 45.46 & 1471 & 7.25 & 2.45 & 0.62 & 0.123 \\
\hline
\Sdd & 9064 & 44.64 & 2156 & 10.62 & 2.41 & 0.57 & 0.135 \\
\hline \hline
\end{tabular}
\end{small}
\end{table}

In Tab. \ref{tab:remnant} we show the properties of the remnants
obtained from the R3 resolution, since due to the high computational costs the R4 resolutions
are not evolved for a long time after the merger. Until the end of our simulations all
the runs except \Suu~collapsed into a black hole, see Fig.~\ref{fig:grhd_rho_max}
where the maximum of the density is shown as an indicator of the BH formation.
In general, the lifetime of the HMNS decreases when we go from the aligned spin setups to the
antialigned spin setups, an indicator that the presence of spins
influences the angular momentum support counteracting the gravitational collapse,
see also~\cite{Kastaun:2014fna,Dietrich:2016lyp}.
Aligned spin configurations, and~\Suu~in
particular, have additional angular momentum support which allows a
longer HMNS lifetime. Similar behavior was also found in
\cite{Dietrich:2016lyp}. While we find that aligned spin configurations lead to more
massive disks and less massive BHs, cf.~\cite{Kastaun:2013mv,Bernuzzi:2013rza},
which is directly caused by the delayed BH formation 
which allows for better angular momentum and matter redistribution into the outer 
layer of the remnant, we do not find any trend in the remnant spins. This can be
attributed to the fact that more refinement is required to resolve the BH formed after
the merger and therefore the inferred properties can incur some errors.

\section{Ejecta and Kick estimates}
\label{sec:ejectaandkicks}

\subsection{Ejecta}
\label{subsec:ejecta}

During our simulations, unbound matter is mainly 
ejected in the very late inspiral from the tidal tail ejection
mechanism or from shock heating during the collision
of the cores of the NSs. In general, our simulations are too short 
to estimate properly disk wind ejecta. 

We compute the amount of ejected matter as
shown for the R3 and R4 resolution simulations
in~Tab.~\ref{tab:ejecta}.
In general, we mark matter as unbound if it fulfills
\begin{equation}
\label{eq:unbound}
u_t<-1 \quad \text{and} \quad v^i x_i >0 \ ,
\end{equation}
where $u_t = -W (\alpha - \beta_i v^i)$
is the time component of the fluid 4-velocity (with a lowered index),
$\alpha$ is the lapse, $\beta^i$ is the shift vector,
$W$ is the Lorentz factor, and $x^i = (x,y,z)$.
For Eq.~\eqref{eq:unbound} we assume that the fluid elements follow geodesics
and require that the orbit is unbound and has an outward pointing velocity,
cf.~also~\cite{East:2012ww}.

\begin{figure*}[tp]
\centering
\includegraphics[width=\linewidth]{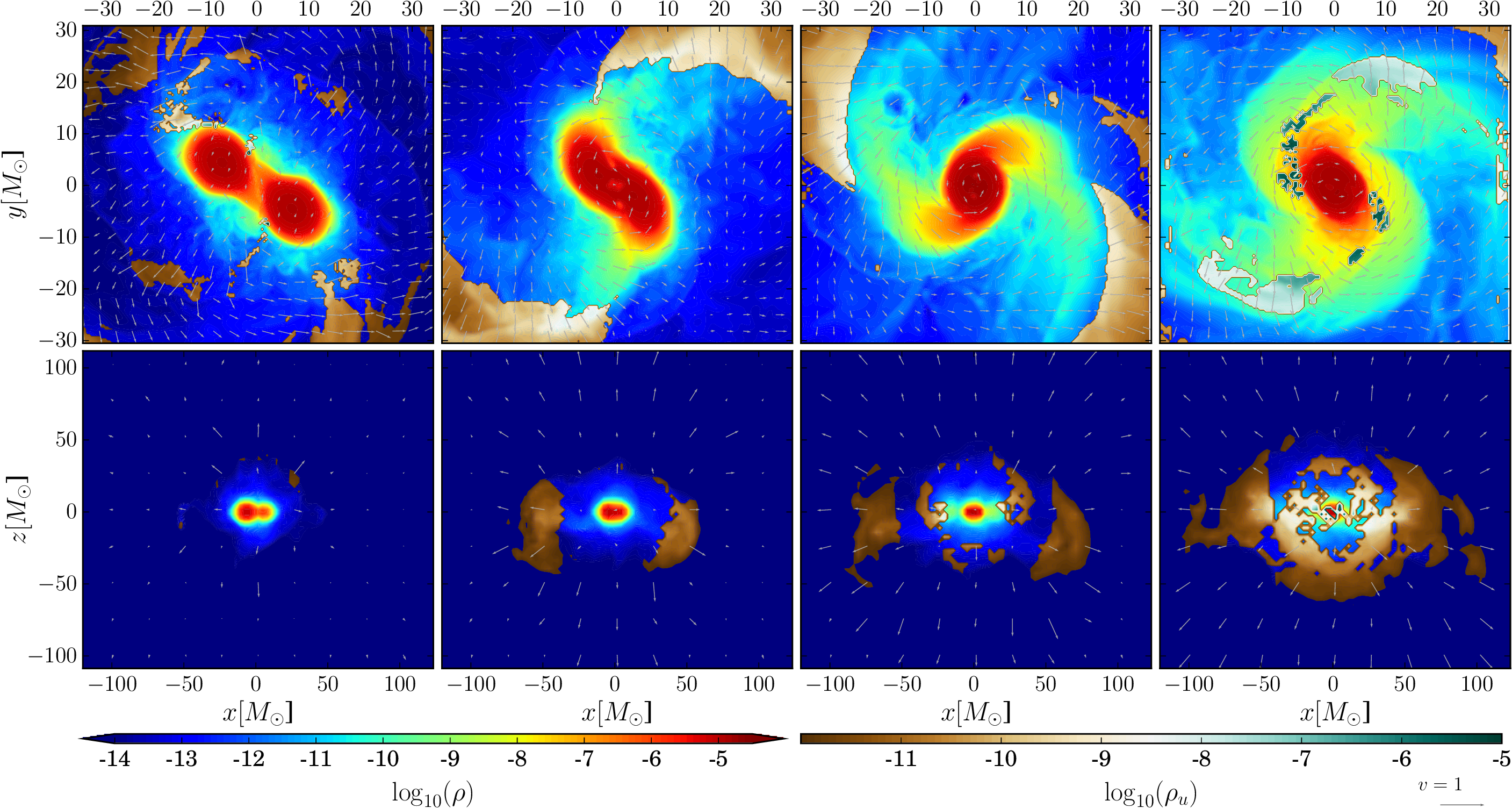}
\caption{
2D-plots for the~\Ssw~configuration showing
the density and velocity field at different times close to the merger, with the unbound
material shown in the brown to green color scale, while the bound material is
shown in a blue to red color scale. \textit{Top row:} plots show
the $xy-$plane covering a distance of $\sim 88$ km in each direction;
\textit{Bottom row:} plots show the $xz-$plane, where each
direction is covering a distance of $\sim 293$ km.
\textit{Columns one to three}: Time snapshots when the surfaces of the stars touch 
until the cores of the NSs finally merged. 
\textit{Fourth column}: Postmerger phase when a hypermassive NS
has been formed. Interestingly, we see that
in the final phase of the merger unbound matter
is ejected asymmetrically due to the ``bobbing'' motion
that this system undergoes. Such an asymmetrical matter
ejection is capable of imparting a kick velocity
to the merger remnant.}
\label{fig:2dejecta}
\end{figure*}

Bound and unbound matter along with
their velocity profile is shown for
the~\Ssw~case in Fig.~\ref{fig:2dejecta}.
Here, we see that the matter ejection does not happen until
the NSs collide (column one). After that (column two),
unbound matter characterized with a density $\sim \mathcal{O}(10^ {-9}) - \mathcal{O}(10^ {-8})$
($\sim \mathcal{O}(10^8) - \mathcal{O}(10^9)$ g cm$^{-3}$)
can be seen coming out from the tidal tail mostly in the orbital
plane (note that this case shows a ``bobbing'' motion of the orbital plane,
see Fig.~\ref{fig:dynafigs}). These ejecta quickly expand
into the volume surrounding the system, dropping
in density by several orders of magnitude. Once the cores of the NSs
have merged (column three) there are also ejecta in the direction
normal to the orbital plane due to shock heating. During these
last phases in the merger unbound matter characterized with a density
$\sim \mathcal{O}(10^ {-8}) - \mathcal{O}(10^ {-6})$
($\sim \mathcal{O}(10^9) - \mathcal{O}(10^{11})$ g cm$^{-3}$)
is ejected. In principle,
unlike equal-mass nonprecessing quasicircular BNSs where the matter should be
symmetrically ejected, similar setups for precessing BNSs can
eject matter asymmetrically due to the ``wobbling'' or the ``bobbing''
motion of the system. 
This asymmetrical ejection of matter
would then give rise to electromagnetic counterparts with a
more complicated geometry.

From Tab.~\ref{tab:ejecta} we see that the amount of unbound matter
increases when the spin of the NS is effectively antialigned
to the orbital angular momentum. This indicates that the ejecta is
dominated via shock heating during the merger of
the cores of the two stars, see also~\cite{Kastaun:2016elu,Most:2019pac}.
Overall, $\sim \mathcal{O}(10^ {-3}) - \mathcal{O}(10^ {-2}) \ M_{\odot}$
($\sim \mathcal{O}(10^{30}) - \mathcal{O}(10^{31})$ g)
of unbound matter is ejected for the studied configurations.
We find the relative error in the estimate of the ejecta mass to be $\sim 2\%-40\%$
between the R3 and R4 resolution setups. No strong effect of precession is
found on the ejecta mass within our simulations.

\begin{table}[tp]
\caption{Ejecta mass from the volume integral $M_\text{ej}^{\mathcal{V}}$ (cf.~\cite{Chaurasia:2018zhg})
for the R3 and R4 resolution setups.}
\centering
\begin{tabular}{c|cc}
\toprule
Name & \multicolumn{2}{c}{$M_\text{ej}^{\mathcal{V}} [M_{\odot}]$} \\ 
     & \begin{footnotesize} R3  \end{footnotesize} & \begin{footnotesize} R4  \end{footnotesize}  \\
\hline
\Suu      \quad & \quad $0.0053$ & $0.0043$ \\ 
\Snw      \quad & \quad $0.0045$ &  $0.0062$ \\ 
\Snee     \quad & \quad $0.0031$ &  $0.0054$ \\ 
\Swe      \quad & \quad $0.0111$ &  $0.0162$ \\ 
\Ssw      \quad & \quad $0.0192$ &  $0.0188$ \\ 
\Ssee     \quad & \quad $0.0210$ &  $0.0189$ \\ 
\Sdd      \quad & \quad $0.0275$ &  $0.0192$ \\
\hline \hline
\end{tabular}
\label{tab:ejecta}
\end{table}

\subsection{Kick estimates}
\label{subsec:kicks}

In addition to the kicks obtained from the asymmetrical matter ejection
mechanism briefly
described in the previous subsection,
the anisotropic loss of linear momentum
radiated away via the emission of GWs also
imparts a recoil or kick on the remaining system which
then moves relative to its original center-of-mass frame.
This effect can be particularly pronounced for the inspiral and
merger of two compact objects, for BBH cases;
see e.g.~\cite{Gonzalez:2007hi,Bruegmann:2008bhk,Gonzalez:2006md}.

\begin{figure}
\centering
\includegraphics[width=0.5\textwidth]{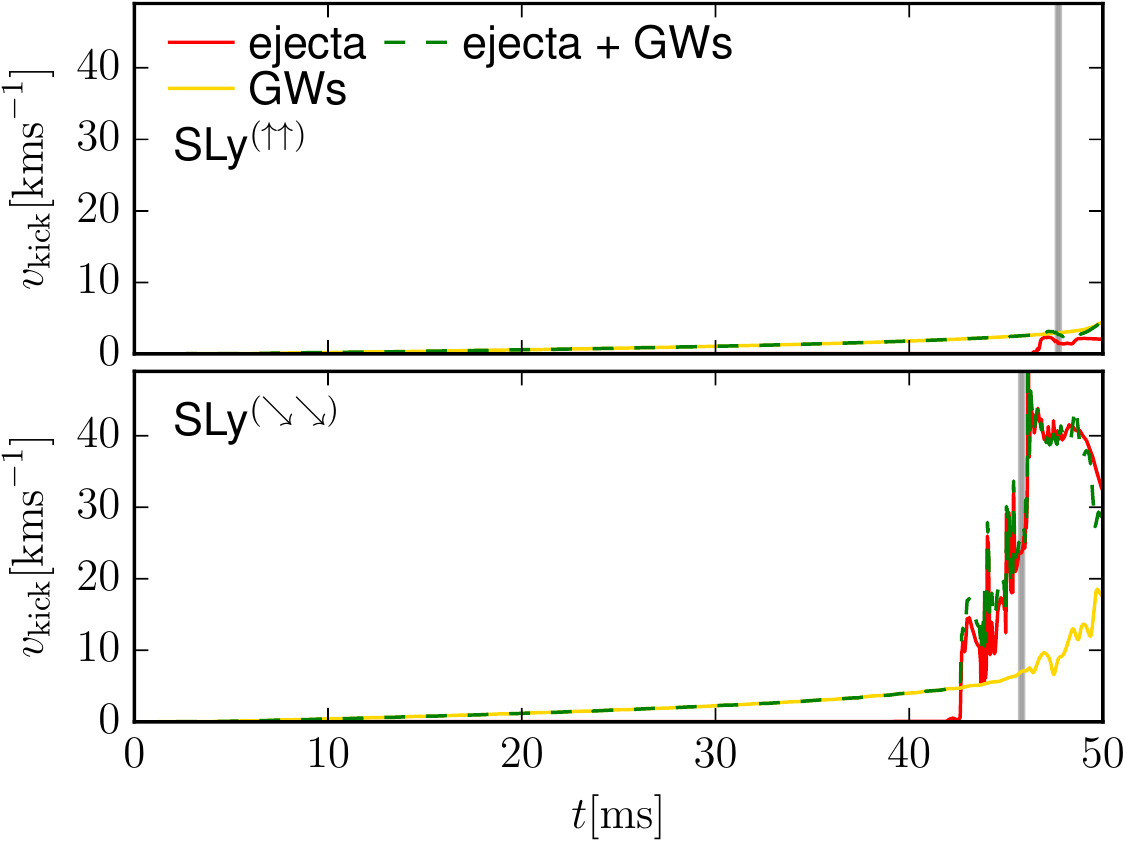}
\caption{Top panel: Kick estimates for the~\Suu~case. The aligned/antialigned
cases do not show the ``bobbing'' or the ``wobbling'' motion of the orbital
plane. Bottom panel: Kick estimates for the~\Ssee~case that shows the
``wobbling'' motion of the orbital plane. The kicks are estimated using the recoil
from the ejecta and the GWs and are shown for the R4 setup. The merger time,
corresponding to the
peak in the (2,2)-mode of GW strain is shown as `gray' line.}
\label{fig:kicks}
\end{figure}

In Fig.~\ref{fig:kicks} we show the estimates for the kick
speed computed from the ejecta and from the emission
of GWs for the R4 resolution setups. The kick estimates
from the ejecta are computed
from the conservation of linear momentum for the
unbound matter whereas the estimates from GWs are computed
using the linear momentum conservation for the GWs using
the relations given in Appendix~\ref{app:EJ_formulas}.
As expected, aligned (and antialigned) systems considered
in this article being symmetrical, the kicks imparted
from the GWs are negligible ($< 5$ km s$^{-1}$).
Furthermore, for the symmetrically misaligned configurations
considered that undergo ``bobbing'' motion, we find the kick
speeds to be in the range $\sim 15-50$ km s$^{-1}$ and
is mostly contributed from the motion of the orbital plane
giving rise to asymmetrical matter ejection. For the asymmetrically
misaligned configurations, for example in the bottom panel
of Fig.~\ref{fig:kicks}, we find that the kicks are again
mostly contributed from the matter ejection, e.g., $\sim 40$ km s$^{-1}$
for the~\Ssee~case. In general, we obtain larger recoils for the
effectively antialigned configurations than for the aligned spin
configurations, but do not see a noticeable difference between the ``wobbling'' and 
``bobbing'' setups. The kicks from the R3 setups for the configurations
shown in Fig.~\ref{fig:kicks} are estimated to be $< 10$ km s$^{-1}$
for the aligned case and $\sim 52$ km s$^{-1}$ for the asymmetrically
misaligned case.

Overall, for all simulated cases the kick imparted from the GW emission contributes 
less than the recoil from unbound matter ejection. 
This might be due to the ``smaller'' spins of neutron stars in comparison to BHs, 
for the latter much larger kicks of 
$\sim \mathcal{O}(10^3)$ km s$^{-1}$, e.g.,~\cite{Gonzalez:2007hi},
can be obtained due to the anisotropic emission of GWs.

\section{Gravitational Waves}
\label{sec:GWs}

\begin{figure*}[tp]
\centering
\includegraphics[scale=0.35]{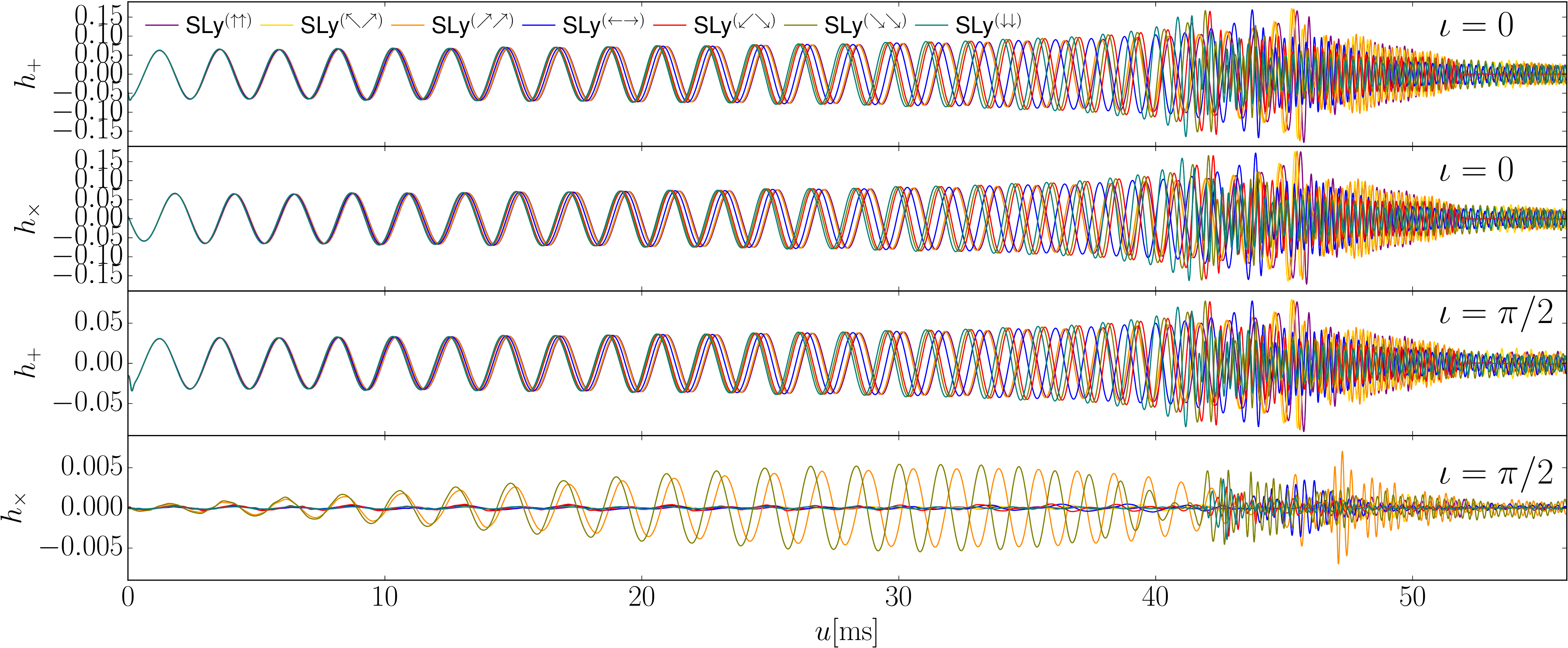}
\caption{Gravitational wave strains $h _+$ (first and third panels)
and $h _{\times}$ (second and fourth panels) for the inclinations
$\iota = 0$ (face on, top panels) and $\iota = \pi/2$ (edge on, bottom panels).}
\label{fig:GW_strain}
\end{figure*}

\subsection{Qualitative discussion}

The individual modes with respect to the $-2$-spin-weighted spherical harmonics of the
curvature and the metric scalars are obtained following 
Sec.~VIA of~\cite{Chaurasia:2018zhg} and references therein.
Additionally, in this article we compute the GW strain $h$ by summing
all modes up to $\ell \leq 4$. All waveforms are shown against the
retarded time
\begin{align}
u = t - r _* = t - t _{\rm extr.} - 2M\ln (r _{\rm extr.}/2M - 1).
\end{align}
Figure~\ref{fig:GW_strain} shows the $h _+$ and $h _{\times}$
polarizations of the GW strain,
\begin{align}
h _+ - ih _{\times} = \sum _{\ell =2} ^4 \sum _{m =-\ell} ^\ell h _{\ell m}
\ ^{-2} Y _{\ell m} (\theta=\iota, \phi=0),
\end{align}
for two inclinations: face on $\iota = 0$ (two top panels) and
edge on $\iota = \pi/2$ (two bottom panels). Similar
inferences can be made as those from column-three of
Fig.~\ref{fig:dynafigs}. 
As expected, we see that for $\iota = 0$ (face on) any imprint of precession
is hardly visible and that the $h _+$ or $h _{\times}-$polarizations 
have the same magnitude. The GW strain is, as discussed before, mainly determined by the effective
spin $\chi_{\rm eff}$ and the spin-orbit-contribution.

Precession effects with more than one precession cycle are visible in
$h _{\times}$ for $\iota = \pi/2$ (edge on) for~\Snee~\&~\Ssee. For these cases, the amplitude of
$h _{\times}(\iota=\pi/2)$ is about $10$ times smaller than for $h _+(\iota=\pi/2)$ 
and 30 times smaller than $h _{\times}(\iota=0)$ or $h _{+}(\iota=0)$. 
For the nonprecessing cases, the signal amplitude of $h _{\times}$ is
even smaller as already seen in
Fig.~\ref{fig:dynafigs} for the (2,1)-mode of the GW strain.

\begin{figure}
\centering
\includegraphics[width=0.5\textwidth]{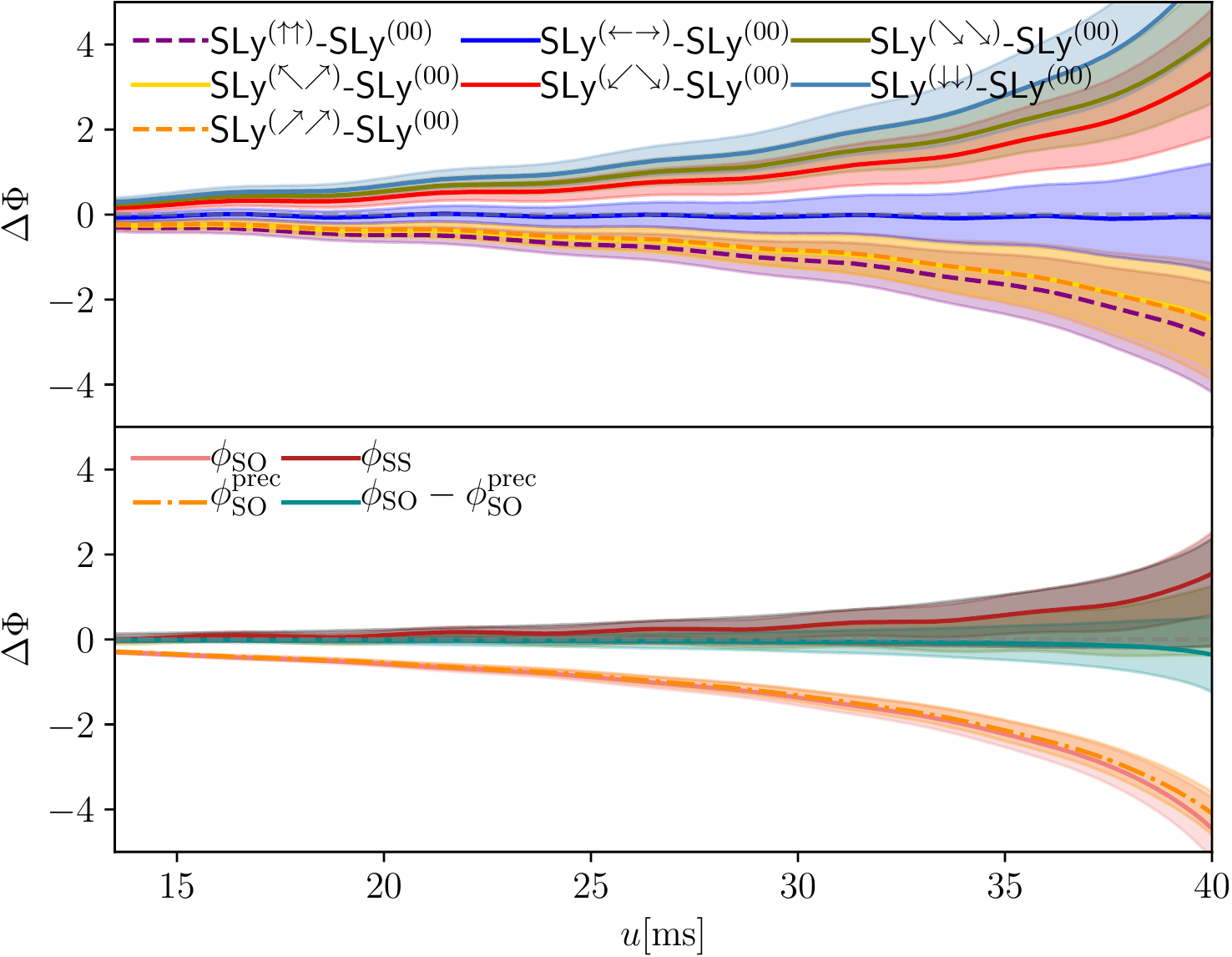}
\caption{Top panel: Phase differences for all
spinning configurations with respect to the
irrotational case for the (2,2)-mode of the GW strain.
The errors represented by the shaded regions are estimated by computing
the phase differences for
different resolutions. Note again that the irrotational case
data used, namely R3 (0.118 $M _\odot$) and R4 (0.078 $M _\odot$) resolutions,
are not of exactly the same resolution as the other configurations simulated and
therefore the error estimates should be taken as conservative estimates.
Bottom panel: Estimate of spin-orbit and spin-spin contribution to the
phase from the aligned/antialigned configurations
as described in the text.}
\label{fig:dephasing1}
\end{figure}

\subsection{Phasing analysis}

In this subsection we discuss briefly the phase evolution
for the different configurations by considering the phase
differences between them for the (2,2)-mode of the
GW strain $h$. Note that the irrotational case~\Soo~is
aligned with~\Swe~configuration in the interval
$\hat{\omega} := M\omega _{22} \in [0.040,0.048]$
for the analysis purpose.

In Fig.~\ref{fig:dephasing1}, the phase differences are shown
for the spinning configurations with respect to the nonspinning
configuration (top panel). It is clearly visible that the 
effectively antialigned systems undergo
accelerated inspiral and the aligned systems
undergo decelerated inspiral. 
These phase differences are again dominated by the
leading-order spin-orbit coupling. The irrotational case and
the~\Swe~case are almost indistinguishable with negligible difference
with respect to phase difference. 

To isolate the effect of different contributions to the
phase evolution we consider, similar to the binding energy discussion,
different linear combinations of the numerical simulations, 
but we emphasize that this analysis is not gauge invariant, i.e., 
it only allows for a qualitative interpretation.
In particular, we consider for the spin-orbit contribution,
\begin{equation}
\phi _\text{SO} = \frac{\phi [\text{SLy}^{(\uparrow \uparrow)}] -
\phi [ \text{SLy}^{(\downarrow \downarrow)}]}{2},
\end{equation}
and for the spin-spin contribution,
\begin{equation}
\phi _\text{SS} = \frac{ \phi [\text{SLy}^{(\uparrow \uparrow)}] +
\phi [\text{SLy}^{(\downarrow \downarrow)}] }{2} -
\phi [\text{SLy}^{(00)}]. 
\end{equation}

To estimate the effect of precession, we also compute 
\begin{equation}
\phi ^\text{prec} _\text{SO} = \sqrt{2} \frac{\phi [\text{SLy}^{(\nwarrow \nearrow)}] -
\phi [ \text{SLy}^{(\swarrow \searrow)}]}{2},
\end{equation}
where the factor $\sqrt{2}$ is introduced to compensate for the fact that the effective spin 
of the precessing configurations is smaller than for the spin-aligned setups. 

Figure~\ref{fig:dephasing1} (bottom panel) shows these
contributions. 
Considering the spin-orbit contribution, we find almost no difference between 
the spin-aligned and the precessing setups, in fact, the difference between both 
contributions can not be resolved with our simulations; 
cf.~solid green line in the bottom panel of Fig.~\ref{fig:dephasing1}
and the discussion on waveform accuracy in Appendix~\ref{app:accuracy}.
Overall, the spin-orbit contribution
dominates so that the spin-spin effect is about a factor 3 smaller. 
Considering our error estimate, we find that for the last few orbits, 
the spin-spin contribution is reliably measured as nonzero.

\begin{figure}
\centering
\includegraphics[width=0.5\textwidth]{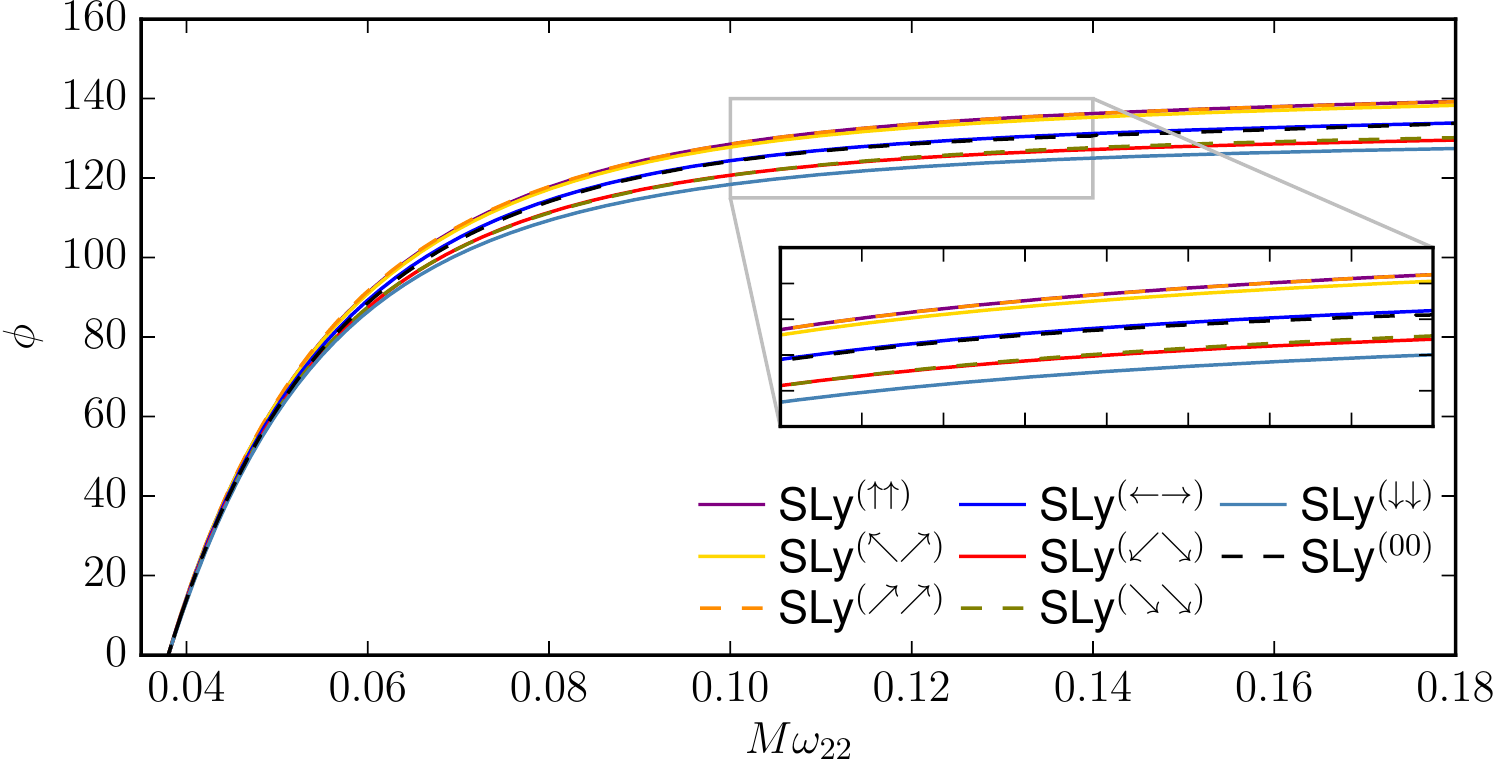}
\caption{$\phi (\hat{\omega})$ accumulated in $\hat{\omega}
\in$ [0.038,0.18] for all the configurations considered
for the R4 resolution.}
\label{fig:dephasing2}
\end{figure}

In order for a more quantitative analysis, 
we analyze the phasing of the waves by considering $\phi(\hat{\omega})$. 
We fit $\phi(\hat{\omega})$ with a function,
\begin{align}
f(\hat{\omega}) = \frac{\sum _{n=0} ^{4} a_n \hat{\omega}^n }{\sum _{n=0} ^{4} b_n \hat{\omega}^n},
\end{align}
eliminating this way the residual eccentricity oscillations in the NR data. We then
align the curves to start at the same frequency $\hat{\omega} = 0.038$.
The phase comparison is restricted to the frequency interval
$\hat{\omega} = [0.038,0.18]$ which corresponds to physical GW frequencies
$\sim$ 455 - 2153 Hz. Figure~\ref{fig:dephasing2} summarizes our results of
the comparison of the accumulated phase difference in the mentioned frequency
interval. Overall, we again find the dominant spin-orbit contribution to give rise to the
different accumulated phases at a particular frequency for the different configurations.
Precession effects are again hardly visible. One can also see in the inset plot
in Fig.~\ref{fig:dephasing2} that the~\Soo~and~\Swe~are indistinguishable considering
the accumulated phases for the dominant GW mode.

\subsection{Comparison with precessing tidal GW approximant}

\begin{figure*}[tp]
\centering
\includegraphics[scale=0.30]{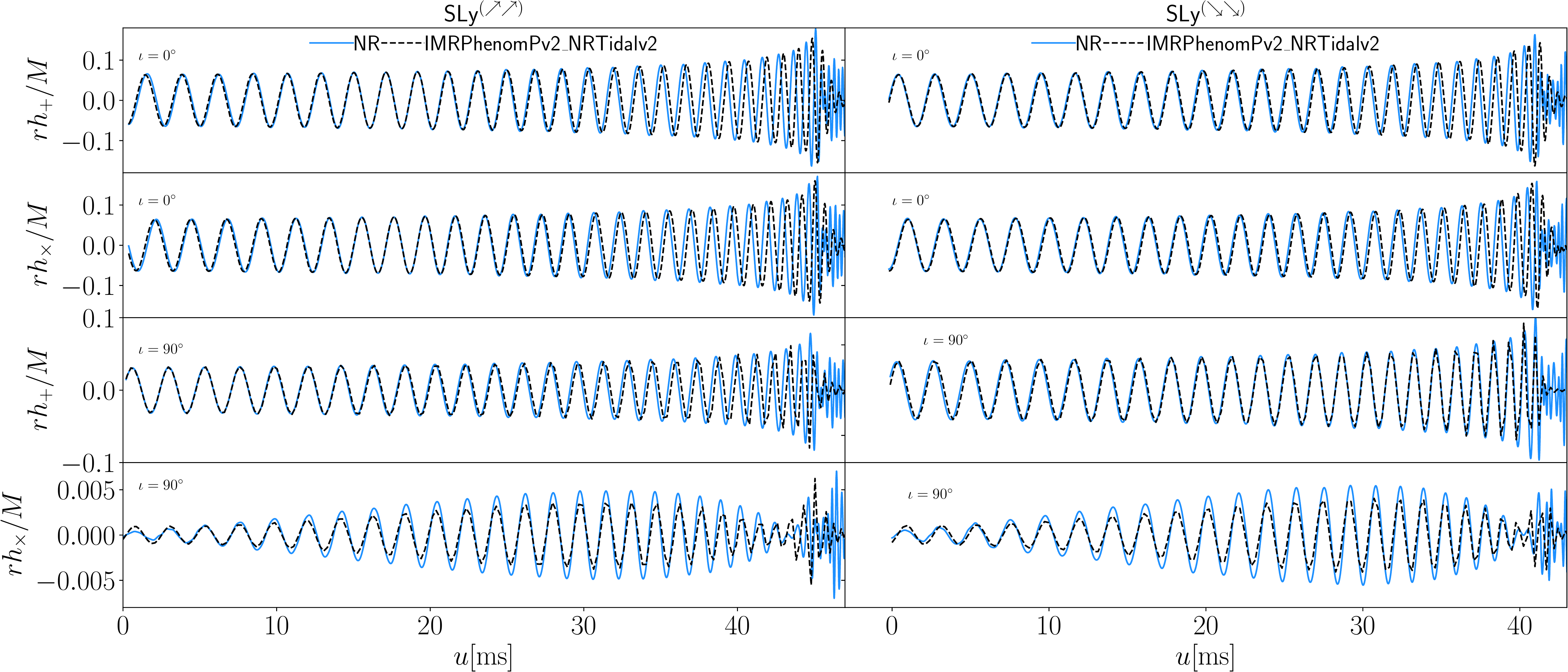}
\caption{Rescaled GW strains for the precessing systems~\Snee~(left panel)
and~\Ssee~(right panel) (blue, solid curves) for the R4 resolution
compared with the \texttt{ IMRPhenomPv2\_NRTidalv2} model (black, dashed curve).
Results for $h _+$ are shown in first and third panel and for $h _{\times}$
are shown in the second and fourth panels for the inclinations $\iota = 0$ (face on)
in top panels and $\iota = \pi/2$ (edge on) in bottom panels. Note that
we find small differences in the amplitudes (as visible in the $h _{\times}$-panels
for the $\iota = \pi/2$ case)
indicating the importance of future GW waveform model development.}
\label{fig:align_strain}
\end{figure*}

One important advantage of full numerical relativity simulations 
is their potential usage for the validation of existing waveform approximants.
Until now, the only two existing precessing, tidal waveform models 
are \texttt{ IMRPhenomPv2\_NRTidal}~\cite{Dietrich:2018uni} and 
\texttt{IMRPhenomPv2\_NRTidalv2}~\cite{Dietrich:2019kaq}.
We focus on the comparison against \texttt{IMRPhenomPv2\_NRTidalv2} in the following. 

\texttt{ IMRPhenomPv2\_NRTidalv2} is a phenomenological, frequency domain, 
tidal-precessing model that augments the aligned-spin binary black hole model, 
\texttt{IMRPhenomD}~\cite{Khan:2015jqa,Husa:2015iqa} 
with the \texttt{NRTidalv2}~\cite{Dietrich:2019kaq} tidal description.
In addition, it incorporates all the relevant EOS dependent spin-spin effects and 
cubic-in-spin effects at 2PN, 3PN, and 3.5PN; and in addition a tidal amplitude 
correction that is added to the binary black hole amplitude. 
To ensure that the system can describe precession effects, 
the aligned-spin waveform is modified by following the 
framework outlined in~\cite{Schmidt:2012rh,Schmidt:2014iyl}.  

We align the \texttt{IMRPhenomPv2\_NRTidalv2} waveforms with the numerical relativity waveforms 
by varying the time translations and phase shifts. 
To obtain the phase and time shift, we minimize the phase difference between the waveforms in 
the time interval $u \in [5, 18]$ ms which corresponds to roughly $8$ GW cycles. 
In addition, we also vary slightly the ``reference frequency'' at which the orientation 
of individual spins are fixed for the construction of the precessing \texttt{IMRPhenomPv2\_NRTidalv2} model. 
While the numerical relativity simulations have an initial frequency of $\sim 407~\rm Hz$, 
we use $410~\rm Hz$ instead to account for the initial transition caused by gauge changes 
in the simulation.

The comparison among the precessing systems~\Snee~(left panel) and~\Ssee~(right panel) and the
\texttt{ IMRPhenomPv2\_NRTidalv2} model is shown in Fig.~\ref{fig:align_strain} for two inclination
angles, $\iota = 0$ (face on) in top panels and 
$\iota = \pi/2$ (edge on) in bottom panels. 
We find that the model is in good agreement with the NR waveforms and also captures precessing motion, i.e., the modulation 
of the GW strain, adequately as shown in bottom panels. 
The phase difference between the numerical relativity waveforms and 
\texttt{ IMRPhenomPv2\_NRTidalv2} is about 1 radian for an inclination of $\iota = 0$ and about 
1.2 radian for $\iota=\pi/2$, just before the merger.

\subsection{Postmerger}

\begin{figure}
\centering
\includegraphics[width=0.5\textwidth]{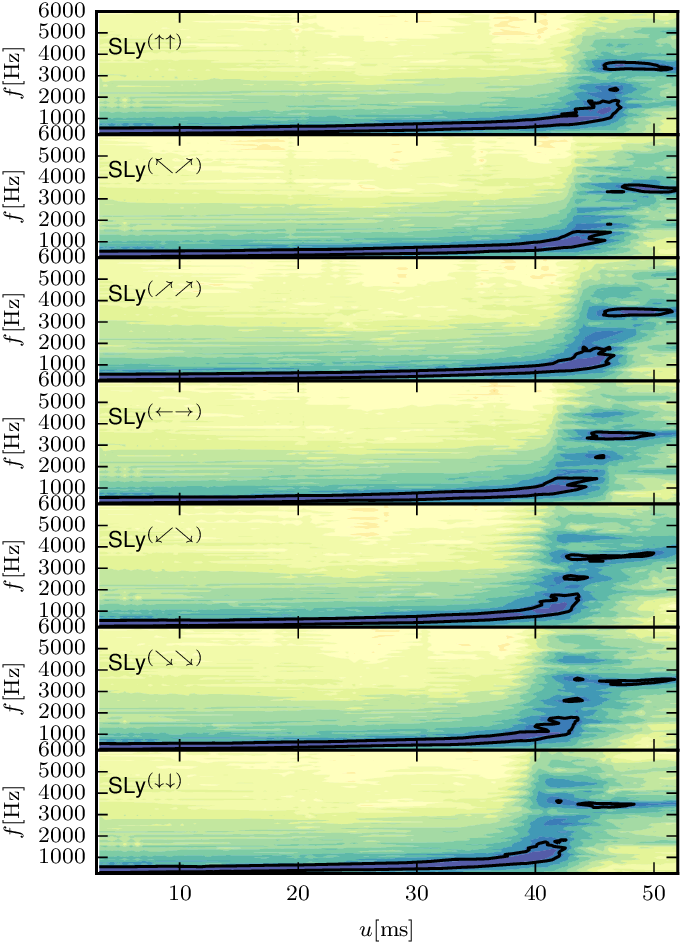}
\caption{Spectrograms and the corresponding contours for all the configurations
computed using the GW strain $h$.
An inclination angle of $\iota = \pi/2$ is assumed for all the plots and a
logarithmic color scale is used. All individual chunks of the spectrogram have a
length of $\sim$ 2 ms and a tapering with a tanh-function is
applied before Fourier transforming to minimize oscillations.}
\label{fig:spectrogram}
\end{figure}

To understand the postmerger evolution of the GW signal
we compute the spectrograms as described
in~\cite{Chaurasia:2018zhg}. Figure~\ref{fig:spectrogram}
shows the spectrograms for all the configurations under
the assumption of $\iota = \pi/2$. 
In Tab.~\ref{tab:post-merger} we report important characteristic frequencies, namely, 
the merger frequency and the postmerger frequencies $f _1$,
the dominant $f _2$ frequency, and $f _3$.

We find that for our chosen EOS and masses the dominant $f _2$-peak frequency 
lies at $\approx 3400$ Hz.
In addition to the $f _2$-peak frequency other side peaks
and frequencies are visible. These peaks are harmonics
of the $f _2$ frequency and have amplitudes that are
typically $2$ to $3$ orders of magnitude smaller~\footnote{Note that we follow in 
our notation~\cite{Dietrich:2015pxa} and not~\cite{Takami:2014zpa} 
about the classification of $f_1$ and $f_3$.}.
These peaks correspond to emission at about
$f _1 \approx 1800$ Hz and $f _3 \approx 5600$ Hz,
respectively.

We find that the merger frequencies are higher for the
aligned spin cases than for the antialigned cases; cf.~\cite{dbt_mods_00029293}.
The postmerger frequencies reported in Tab.~\ref{tab:post-merger}
are obtained from the individual modes of the GW strain
and in some cases were not available possibly due to
low signal amplitude or the lifetime of the remnant before BH formation. 
However, in Fig.~\ref{fig:spectrogram}
where the spectrogram was obtained from $h$ those frequencies
are visible albeit with smaller amplitudes relative to the
prominent $f _2$ frequency. 
The frequency estimates have typical uncertainties of $\sim 50-100$ Hz.

For comparison, the estimates for the dominant $f _2$ frequency
using Ref.~\cite[Eq.~(8)]{Tsang_2019} gives a frequency
of $\sim 3372$ Hz and the quasiuniversal relation of 
Ref.~\cite[Eq.~(13)]{Breschi_2019} gives a frequency
of $\sim 3435$ Hz. Both relations do not include spin effects and 
their estimates are below our simulation results, 
but are generally in agreement if the uncertainties of the quasiuniversal
relations and our numerical relativity simulations are taken into account. 
This is interesting and hints towards the fact that while spin affects
the postmerger dynamics, it only has a minor effect on the main postmerger 
emission frequency as outlined in~\cite{Bauswein:2015vxa}. 
Other previous simulations clearly showed spin effects~\cite{Bernuzzi:2013rza}, 
so that we conclude that more simulations focusing specifically on the postmerger evolution 
are needed to solve the existing tension. 

\begin{table}[t]
\caption{
Postmerger properties. The columns give the name of the configuration, 
the dimensionless merger frequency $M\omega_{\rm mrg}$, the dimensionful merger frequency
$f_{\rm mrg}$ (in Hz), and the dominant postmerger frequencies extracted from the
$(2,1)$, $(2,2)$, and $(3,3)$ modes of GW strain $h$. We mark ``$-$''
for cases where the frequencies could not be extracted properly.}
\label{tab:post-merger} 
\centering
\begin{tabular}{cccccc}
\toprule
Name & $M\omega_\text{mrg}$ & $f_{\rm mrg}$ & $f_1$ & $f_2$ & $f_3$ \\ 
     &                      & [Hz]          &  [Hz] &  [Hz] & [Hz] \\
\hline 
\Suu &  $0.165$& $1974$ & $1845$ & $3358$ & $5187$\\
\hline
\Snw &  $0.170$& $2034$ & $1794$ & $3557$ & $5446$\\
\hline
\Snee &  $0.177$& $2118$ & $1826$ & $3620$ & $5351$\\
\hline
\Swe &  $0.150$& $1795$ & $-$ & $3431$ & $5855$\\
\hline
\Ssw &  $0.150$& $1795$ & $-$ & $3400$ & $-$\\
\hline
\Ssee &  $0.143$& $1711$ & $1826$ & $3463$ & $5257$\\
\hline
\Sdd &  $0.140$& $1675$ & $-$ & $3447$ & $5918$ \\
\hline
\end{tabular}
\end{table}

\section{Summary}
\label{sec:summary}

In this article we have continued our systematic study of 
the BNS parameter space where we had previously focused on the 
effect of the mass ratio~\cite{Dietrich:2016hky}, spin~\cite{Dietrich:2016lyp}, 
and eccentricity~\cite{Chaurasia:2018zhg}, now, we investigated the influence 
of the spin orientation. 
For this purpose, we have studied seven different configurations, from which 
two setups have aligned/antialigned spins and five setups have misaligned-spins; 
cf.~Tab.~\ref{tab:config} for the simulation details. 
All configurations are simulated for multiple grid resolutions to provide an estimate 
for the uncertainty of our results; cf.~Tab.~\ref{tab:grid}.

In the following, we want to summarize our main findings: 
\begin{enumerate}[(i)]
 \item Depending on the particular spin configuration, we have systems showing a ``bobbing'' 
 motion of the orbital plane, i.e., an up- and downward movement of the plane, and systems 
 showing a ``wobbling'' motion in which the orbital plane precesses. 
 For ``wobbling'' systems the (2,1)-mode of the GW signal is 
 significantly stronger than for the ``bobbing'' or aligned-spin setups.
 \item Spin-orbit and spin-spin contributions to 
 the binding energy can be extracted from our simulations, 
 but no clear imprint of precession effects is visible in 
 our simulations independent of the spin orientation. 
 \item Only for the ``wobbling'' configurations the emitted GWs carry angular momentum that 
 is not parallel to initial orbital angular momentum; cf. Fig.~\ref{fig:angular_mom_rad}. 
 \item The lifetime of the formed HMNS depends on the effective spin of the system 
 and not on the orientation of the spin, so that systems with positive $\chi_{\rm eff}$ 
 have more angular momentum support at merger and consequently a delayed BH formation in the postmerger stage. 
 In these cases, the disk mass increases while the final BH mass decreases.  
 \item For the precessing systems, mass can be ejected anisotropically and 
 the final remnant can obtain a kick of $\sim 40 \rm km/s$. 
 The anisotropic mass ejection of matter contributes more to the final kick 
 velocity than the anisotropic emission of GWs.
 \item Configurations with antialigned spin create a larger ejection of matter compared to spin-aligned systems.
 \item The precessing and tidal GW approximant \texttt{IMRPhenomPv2\_NRTidalv2} is capable 
 of describing the inspiral signal and capturing the precessing motion of the studied cases.
 \item For the astrophysically motivated cases in which only one star has a non-negligible spin, we expect that there will only be a ``wobbling''  motion of the orbital plane and no ``bobbing'' motion. Additionally, if the spin of the individual star is constant, spin-orbit effects will have a smaller impact and the same will be true for the spin-spin terms as there will only be a self-spin term whereas the spin-spin interaction term will vanish. Moreover, we still expect to see orbital hang-up or speed-up effect but with a smaller effect on the orbital dynamics. Other such inferences based on the presented set of simulation results also follow.
\end{enumerate}

Overall, this work has been a first step towards a better understanding of precession effects 
for BNS systems, but further simulations for unequal-mass systems, unequal spins, and higher spins
need to be studied in the near future. To allow the best usage of our
simulation data, we will release the waveform signals
in the near future as a part of the CoRe database~\cite{CoRe,Dietrich:2018phi}. 

\begin{acknowledgments}
  We thank Sergei Ossokine for helpful discussions. 
  S.~V.~C.~was supported by the DFG Research Training 
  Group 1523/2 ``Quantum and Gravitational Fields'' and
  by the research environment grant "Gravitational Radiation and
  Electromagnetic Astrophysical Transients (GREAT)"
  funded by the Swedish Research council (VR) under Dnr. 2016-06012.
  T.~D.~ acknowledges support by the European Union’s Horizon 2020 research and 
  innovation program under grant agreement No 749145, BNSmergers. 
  M.~U. was supported by Funda\c{c}\~ao de Amparo \`a Pesquisa do Estado 
  de S\~ao Paulo (FAPESP) under process 2017/02139-7.
  B.~B., R.~D., and F.~M.~F.\ were supported in part by DFG grant BR 2176/5-1.
  W.~T.\ was supported by the National Science Foundation
  under grant PHY-1707227.
  Computations were performed on 
  the supercomputer SuperMUC at the LRZ
  (Munich) under the project number pr48pu and pn56zo
  and on the ARA cluster of the University of Jena.
\end{acknowledgments}

\appendix

\section{Radiated Energy, Angular Momentum and Linear Momentum
Computation}
\label{app:EJ_formulas}

To compute the amount of
energy, angular momentum and linear momentum
radiated away from the system in the form of
gravitational radiation we use the
relations as given on pages 313-316
of~\cite{Alcubierre:2008}. The energy is
computed from the time integral of
\begin{align}
\frac{dE}{dt} = {\lim _{r \rightarrow \infty}} \
\frac{r ^2}{16 \pi} \ {\sum _{\ell,m}} \
\left\lvert \ {\int ^t _{-\infty}} A ^{\ell,m} \ dt' \ \right\rvert {^2}.
\end{align}
The angular momentum vector is computed
from the time integral of
\begin{align}
\frac{dJ _x}{dt} = &{-\lim _{r \rightarrow \infty}} \
\frac{i r ^2}{32 \pi} \ {\bf Im} \left\lbrace {\sum _{\ell,m}}
{\int ^t _{-\infty}} {\int ^{t'} _{-\infty}} A ^{\ell,m} \ {dt''} \ dt' \right. \nonumber \\
&\times \left. {\int ^t _{-\infty}}
\left( f _{\ell,m} A ^{* \ell,m+1} +  f _{\ell,-m}
A ^{* \ell,m-1} \right)  \ dt' \right\rbrace, \\
\frac{dJ _y}{dt} = &{-\lim _{r \rightarrow \infty}} \
\frac{r ^2}{32 \pi} \ {\bf Re} \left\lbrace {\sum _{\ell,m}}
{\int ^t _{-\infty}} {\int ^{t'} _{-\infty}} A ^{\ell,m} \ {dt''} \ dt' \right. \nonumber \\
&\times \left. {\int ^t _{-\infty}}
\left( f _{\ell,m} A ^{* \ell,m+1} -  f _{\ell,-m}
A ^{* \ell,m-1} \right) \ dt'\right\rbrace, \\
\frac{dJ _z}{dt} = &{-\lim _{r \rightarrow \infty}} \
\frac{i r ^2}{16 \pi} \ {\bf Im} \left\lbrace {\sum _{\ell,m}} \ m
{\int ^t _{-\infty}} {\int ^{t'} _{-\infty}} A ^{\ell,m} \ {dt''} \ dt' \right. \nonumber \\
&\times \left. {\int ^t _{-\infty}} \ A ^{* \ell,m} \ dt' \right\rbrace,
\end{align}
where, $f _{\ell,m} := \sqrt{(\ell-m)(\ell+m+1)}$
= $\sqrt{\ell(\ell+1)-m(m+1)}$ and ${\bf Im}(a+i b) = i b$
for real $a$ and $b$.

The radiated linear momentum is calculated from the
time integral of
\begin{align}
\frac{d P _+}{dt} &= \lim _{r\rightarrow \infty}
\frac{r ^2}{8 \pi} \sum _{\ell, m} \int ^t _{-\infty}
dt' \ A ^{\ell,m} \nonumber \\
&\times \int ^t _{-\infty} dt' \left( a _{\ell,m} A ^{* \ell,m+1} +
b _{\ell,-m} A ^{* \ell -1,m+1} \right. \nonumber \\
&\left.- b _{\ell+1,m+1} A ^{* \ell+1 ,m+1}  \right) \ ,\\
\frac{d P _z}{dt} &= \lim _{r\rightarrow \infty}
\frac{r ^2}{16 \pi} \sum _{\ell, m} \int ^t _{-\infty}
dt' \ A ^{\ell,m} \nonumber \\
&\times \int ^t _{-\infty} dt' \left( c _{\ell,m} A ^{* \ell,m} +
d _{\ell,m} A ^{* \ell -1,m} \right. \nonumber \\
&\left.+ d _{\ell +1,m} A ^{* \ell +1,m} \right) \ ,
\end{align}
where $P _+ = P _x + i P _y$ and we defined the quantities
\begin{align}
a _{\ell,m} &:= \frac{\sqrt{(\ell - m)(\ell+m+1)}}{\ell(\ell +1)} \ , \nonumber \\
b _{\ell,m} &:= \frac{1}{2l} \sqrt{\frac{(\ell-2)(\ell+2)(\ell+m)(\ell+m-1)}{(2\ell-1)(2\ell+1)}} \ , \nonumber \\
c _{\ell,m} &:= \frac{2m}{\ell (\ell +1)} \ , \nonumber \\
d _{\ell,m} &:= \frac{1}{\ell} \sqrt{\frac{(\ell-2)(\ell+2)(\ell-m)(\ell+m)}{(2\ell-1)(2\ell+1)}} \ .
\end{align}

Note that, ${\int ^t _{-\infty}} \ A ^{\ell,m} \ dt' = \dot{h} _{\ell,m}(t)$ and ${\int ^t _{-\infty}} \ {\int ^{t'} _{-\infty}} \ A ^{\ell,m} \ {dt''} \ dt' = h _{\ell,m}(t)$. A `*' in the above expressions denotes
a complex conjugate. Moreover,
\begin{align}
A ^{\ell,m} &= \left\langle {Y ^{\ell,m} _{-2}}, \Psi _4 \right\rangle \nonumber \\
&= {\int ^{2\pi} _0}{\int ^{\pi} _0} \
\Psi _4 \ {Y} ^{*\ell,m} _{-2} \ \sin\theta \ d\theta d\phi,  
\end{align}
where ${Y} ^{\ell,m} _{-2}$ are the spherical harmonics of spin weight $-2$.


\section{Convergence Study}
\label{app:accuracy}

\begin{figure}
\centering
\includegraphics[width=0.5\textwidth]{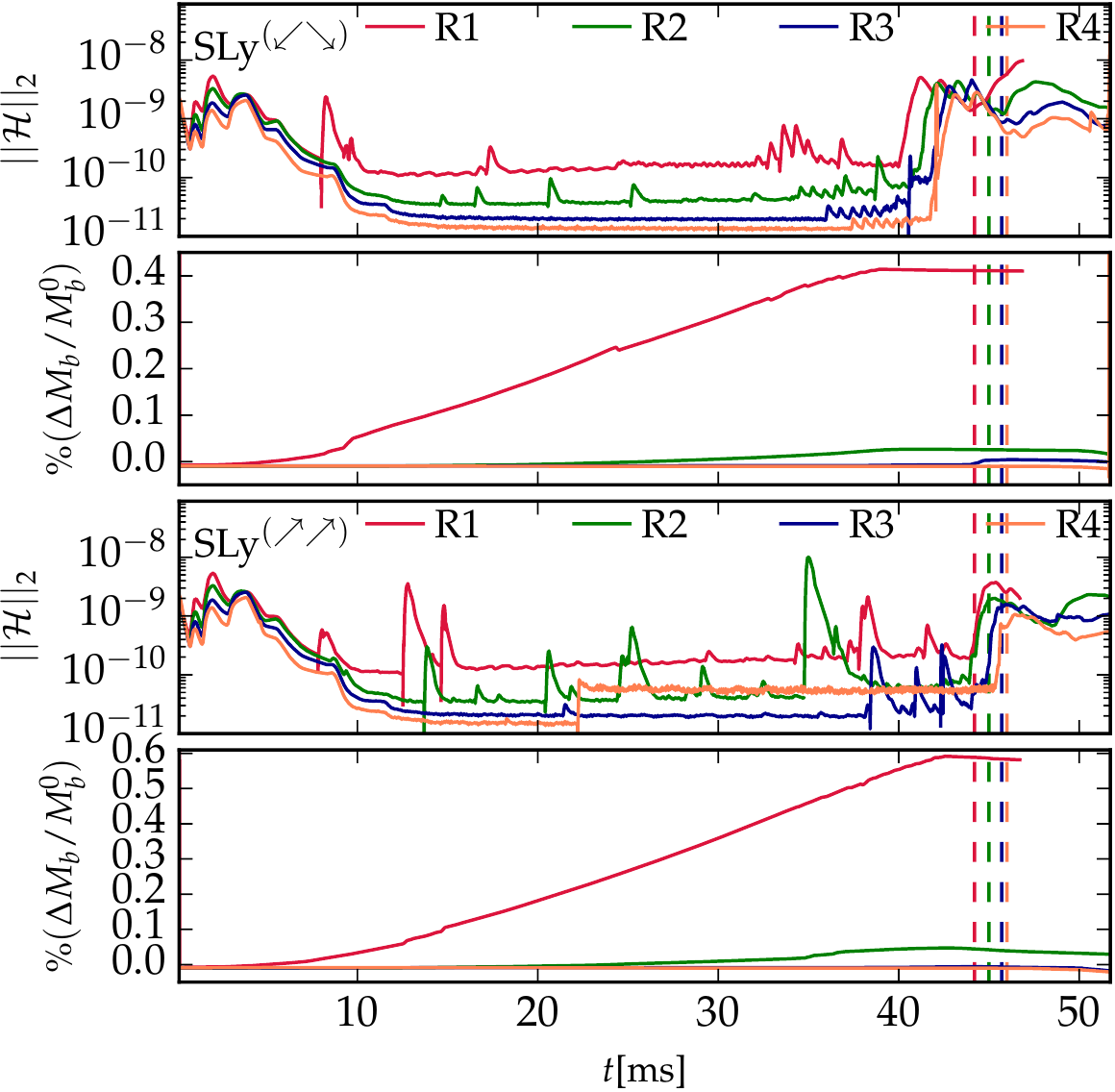}
\caption{Hamiltonian constraint (first and third panel) and
rest mass conservation (second and fourth panel) for the \Ssw case
(top) and the \Snee case (bottom). The merger time, corresponding to the
peak in the (2,2)-mode of GW strain is shown as vertical dashed line for each resolution.}
\label{fig:ham_mass}
\end{figure}

\paragraph*{\textbf{Constraint violation and mass conservation}:-}
For assessing the accuracy and robustness of our
simulations, we present the $L_2$ volume norm of the Hamiltonian constraint
and the conservation of rest mass in
Fig.~\ref{fig:ham_mass} for the \Ssw case
(top panels) and for the
\Snee case (botton panels).

\begin{figure}
\centering
\includegraphics[width=0.5\textwidth]{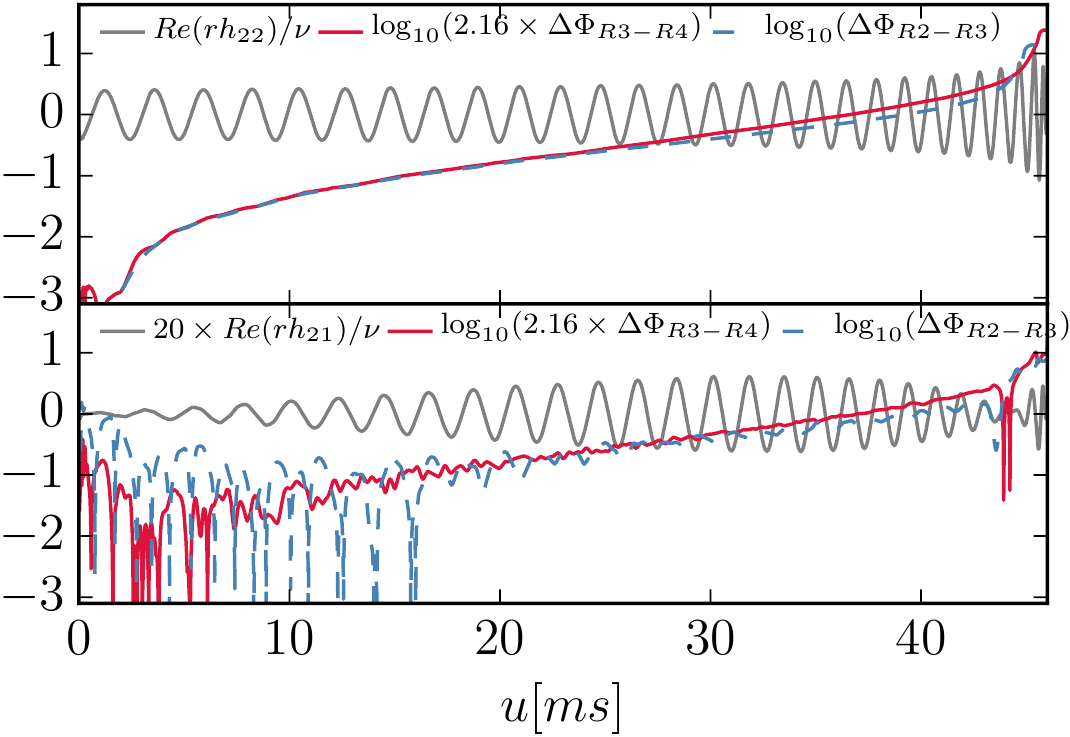}
\caption{Real part of the (2,2) mode (top panel) and (2,1) mode (bottom panel) for
resolution R4 as well as the phase difference between different resolutions for the 
\Snee configuration shown versus retarded time. 
We multiply the amplitude of the (2,1) mode by a
factor of $20$ for better visibility.}
\label{fig:waveform_convg}
\end{figure}

Owing to the constraint propagation and damping
properties of the Z4c
evolution system the constraint stays at or
below the value of the initial data.
Oscillations and spikes in the constraints during the orbital motion, as seen
in~Fig.~\ref{fig:ham_mass}, mainly originate due the inner
refinement levels following the motion of the NSs. After the
merger (vertical dashed lines), those spikes are absent as the stars stay near
the center or move with a very small velocity compared to
during the inspiral phase.
At merger the constraint grows by about two orders of
magnitude due to regridding and to the development
of large gradients in the solution, but it
remains below the initial level. Subsequently,
the violation is again propagated away and damped.
Throughout the simulation we find that the Hamiltonian constraint violation improves monotonically
with increasing resolution for the \Ssw case. For
the \Snee case, only the lower three resolutions, R1, R2 and R3 show this trend whereas for the highest resolution R4,
the constraint violation grows one order of magnitude at $t=~22\rm ms$ during the regridding of the grid, 
but does not decrease afterwards; the exact origin of 
this effect is currently under investigation, 
but it seems that the results presented in the main text are unaffected; 
cf.~also Fig.~\ref{fig:waveform_convg}. 

Violations of rest-mass conservation,
shown in~Fig.~\ref{fig:ham_mass}, happen
at the mesh refinement boundaries and due to the
artificial atmosphere treatment, and possibly due
to mass leaving the computational domain.
From the time evolution of the mass violation,
we find that, independent of the
spin orientation, the resolution R1 shows an
increasing mass during the orbital motion.
This is caused by inadequate resolution
and the artificial atmosphere treatment, see e.g.~\cite{Dietrich:2015iva}. 
For resolutions R2, R3, and R4 the 
rest mass stays constant within $0.1\%$ throughout the
simulation time. The mass loss is caused
by the ejected material which decompresses while it
leaves the central region of the numerical domain. 
Once the density drops by $12$
orders of magnitude,
the material is counted as atmosphere and is not
evolved further.
Consequently, conservation of total mass is violated.
Overall the mass violation is below $0.6\%$ considering all
the resolutions employed in this article.

\paragraph*{\textbf{Waveform accuracy}:-}
In Fig.~\ref{fig:waveform_convg} we present the GW phase difference between different resolutions
for \Snee during the inspiral up to the moment 
of merger, which we define as the time of maximum
amplitude in the (2,2)-mode.
Through the inspiral we see a monotonic
decrease of the phase difference for increasing 
resolution. Note that in
Fig.~\ref{fig:waveform_convg} we have scaled the phase difference from R3-R4 assuming
second-order convergence. We find that the rescaled curve
agrees very well with the R2-R3 curve implying
that our results are in the convergent regime with
increasing resolution.


\bibliography{paper20200820.bbl}

\end{document}